\documentclass[twocolumn,showpacs,showkeys,preprintnumbers,prd,floatfix,nofootinbib,superscriptaddress,letterpaper]{revtex4}
 \usepackage{hyperref}
\usepackage{graphicx}
\usepackage{amsmath}
\usepackage{amssymb}
\usepackage{axodraw}
\usepackage{epsfig}
\usepackage{xcolor}

\def\lsim{\mathrel{\hbox{\rlap{\hbox{\lower4pt\hbox{$\sim$}}}\hbox{$<$}}}}
\def\gsim{\mathrel{\rlap{\lower4pt\hbox{\hskip1pt$\sim$}} \raise1pt\hbox{$>$}}}
\newcommand{\gev}{\ensuremath{\mathrm{GeV}}}
\newcommand{\gevsq}{\ensuremath{\mathrm{GeV^2}}}
\newcommand{\der}{\ensuremath{{\operatorname{d}}}}
\def\msbar{\ensuremath{{\rm{\overline{MS}}}}} 

\begin{document} 

\title{Nuclear PDFs from neutrino deep inelastic 
scattering\footnote{Physical Review D 77, 054013 (2008)}}

\author{
I.~Schienbein,$^{a,b}$\thinspace\footnote{schien@lpsc.in2p3.fr}\ \ 
J.~Y.~Yu,$^a$\thinspace\footnote{yu@physics.smu.edu}\ \ 
C.~Keppel,$^{c,d}$\thinspace\footnote{keppel@jlab.org}\ \ 
J.~G.~Morf\'{\i}n,$^e$\thinspace\footnote{morfin@fnal.gov}\ \ 
F.~Olness,$^{a,g}$ \thinspace\footnote{olness@smu.edu}\ \ 
and
J.F.~Owens$^f$\thinspace\footnote{owens@hep.fsu.edu}\ \ 
}

\affiliation{
$^a$Southern Methodist University, Dallas, TX 75206, USA,
\\
$^b$Laboratoire de Physique Subatomique et de Cosmologie, Universit\'e Joseph Fourier
Grenoble 1, CNRS/IN2P3, Institut National Polytechnique de Grenoble, 
\\
53 Avenue des Martyrs, 38026 Grenoble, France,
\\
$^c$Thomas Jefferson National Accelerator Facility, Newport News, VA 23602, USA,
\\
$^d$Hampton University, Hampton, VA, 23668, USA,
\\
$^e$Fermilab, Batavia, IL 60510, USA,
\\
$^f$Florida State University, Tallahassee, FL 32306-4350, USA
\\
$^g$Theoretical Physics Division, Physics Department, 
CERN, CH 1211 Geneva 23, Switzerland
}

\date{\today}% It is always \today, today,

%   12.38.-t 	Quantum chromodynamics
%   13.15.+g 	Neutrino interactionssuperscriptaddress
%   13.60.-r 	Photon and charged-lepton interactions with hadrons
%   24.85.+p 	Quarks, gluons, and QCD in nuclear reactions
%   
\pacs{12.38.-t,13.15.+g,13.60.-r,24.85.+p\vspace{12mm}}

\begin{abstract}
%\abstract{
% MODIFIED: 12 DEC 2007 
We study nuclear effects in charged current deep inelastic
neutrino-iron scattering in the frame-work of a $\chi^2$ analysis of
parton distribution functions. We extract a set of iron PDFs and show
that under reasonable assumptions it is possible to constrain the
valence, light sea and strange quark distributions.  Our iron PDFs are
used to compute $x_{Bj}$-dependent and $Q^2$-dependent nuclear
correction factors for iron structure functions which are required in
global analyses of free nucleon PDFs.  We compare our results with
nuclear correction factors from neutrino-nucleus scattering models and
correction factors for $\ell^\pm$-iron scattering.  We find that,
except for very high $x_{Bj}$, our correction factors differ in both
shape and magnitude from the correction factors of the models and
charged-lepton scattering.
%}
\end{abstract}


%\preprint{\ldots}
%\\
%\preprint{arXiv:YYMM.NNNN}
\preprint{arXiv:0710.4897}
%\\  \quad
\preprint{LPSC 07-45}
%\\ \quad
\preprint{CERN-PH-TH-2007-199}
%\preprint{hep-ph/yymmnnn}
%\\
%}

\maketitle

%**********************************************************

%%%%%%%%%%%%%%%%%%%%%%%%%%%%%%%%%%%%%%%%%%%%%%%%%%%%%%%%%%%%%%%%%%%%%%%
%   `A2' FIT
%
\def\figAAa{
\begin{figure*}[!t]
\begin{center}
\includegraphics[width=1.80\columnwidth]{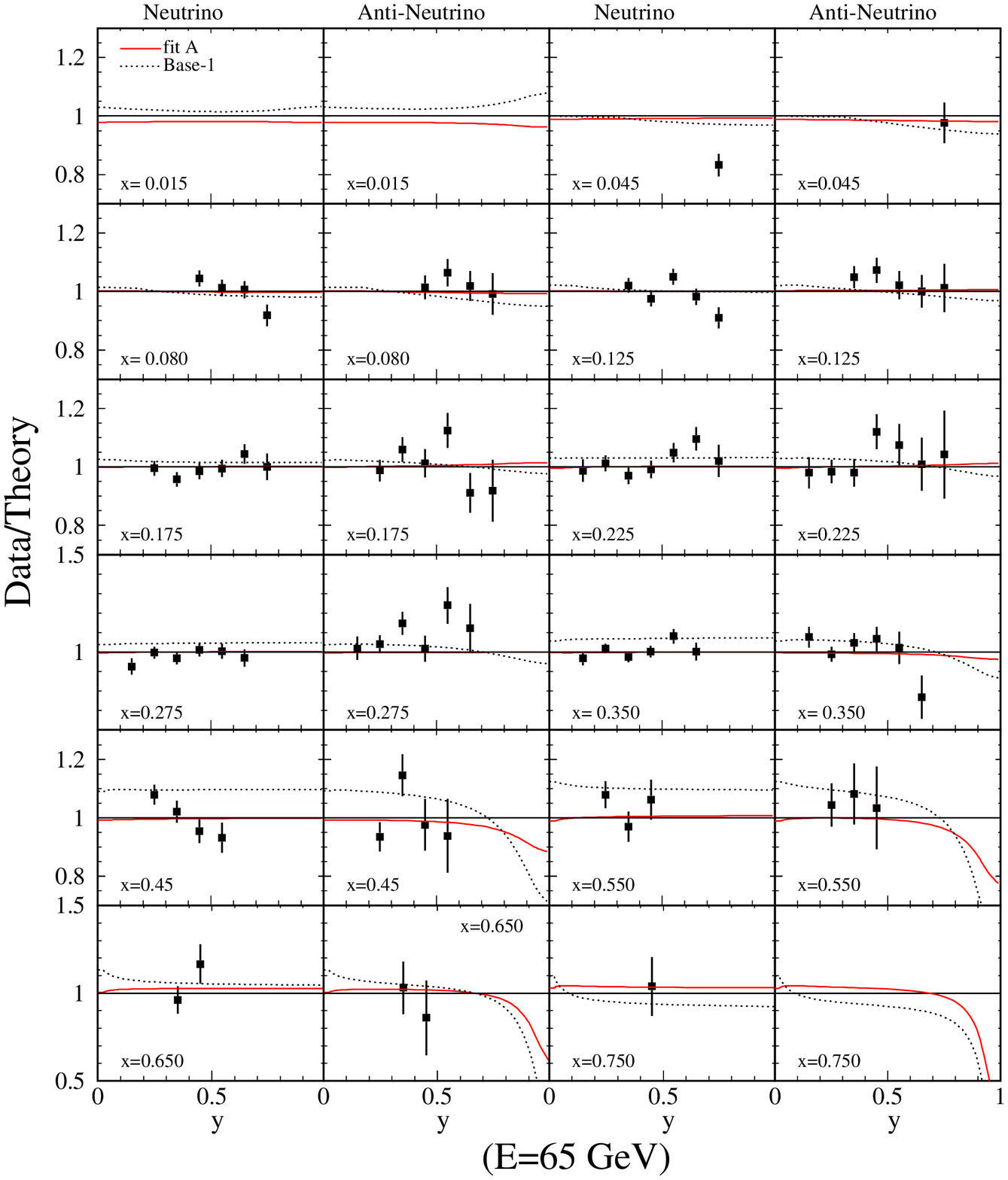} 
%\vspace*{-1cm}
\caption{
Representative comparison of  fit `A2' to the NuTeV 
neutrino and anti-neutrino cross section data.  Shown are the data
points for various $x$-bins versus the inelasticity $y$ for an energy
of $E=65~\gev$ in a data-over-theory representation.
For comparison, we also show results for the Base-1 PDFs (dotted)
and the `A' fit (solid); 
the  fit `A2' imposes  more stringent cuts on $Q>2~\gev$ and $W>3.5~\gev$.
}
\label{fig:fig2a}
\end{center}
\end{figure*}
}

%%%%%%%%%%%%%%%%%%%%%%%%%%%%%%%%%%%%%%%%%%%%%%%%%%%%%%%%%%%%%%%%%%%%%%%
%   `A2' FIT
%
\def\figAAb{
\begin{figure*}[!t]
\begin{center}
\includegraphics[width=1.80\columnwidth]{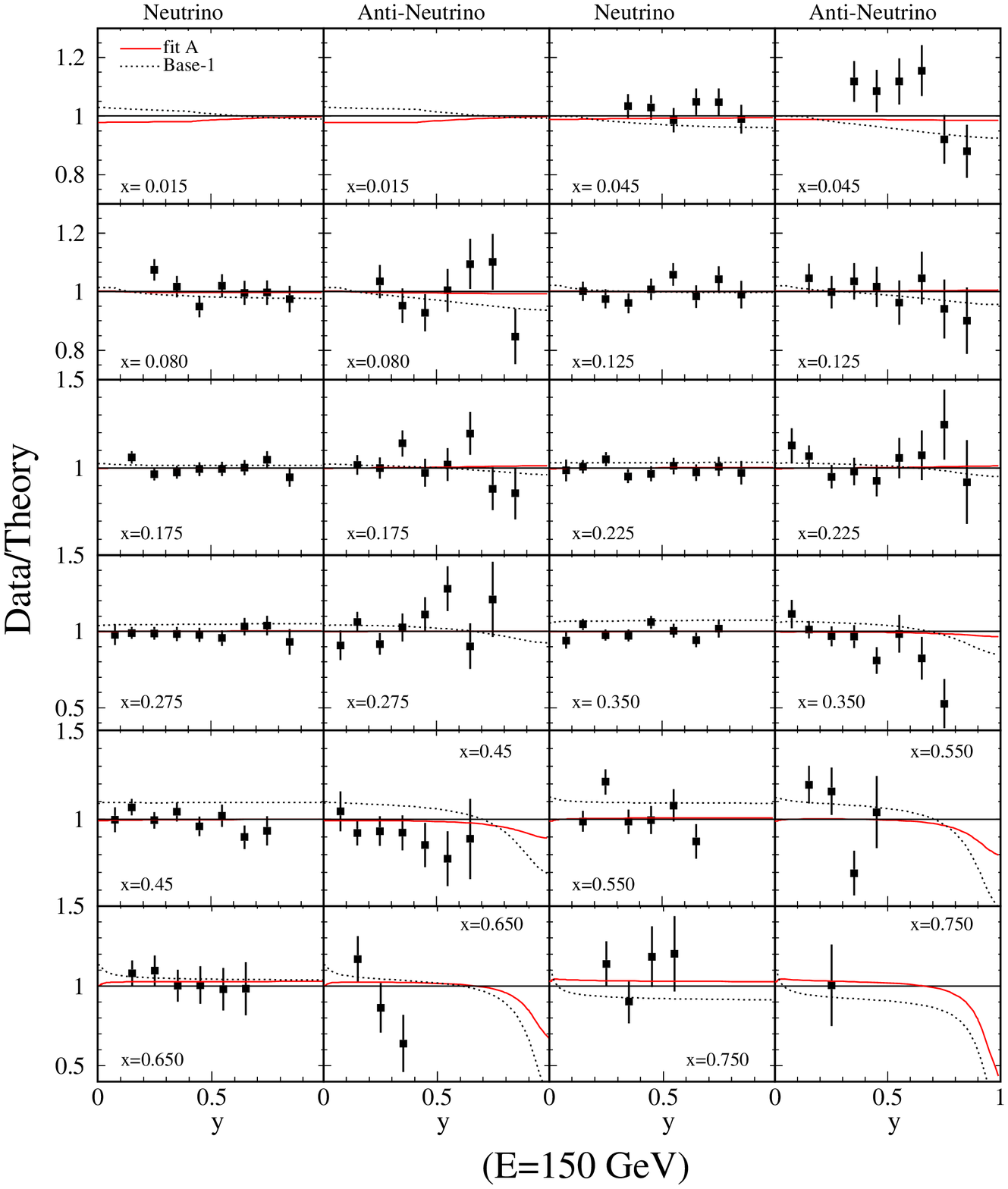} 
%\vspace*{-1cm}
\caption{
The same as in Fig.~\protect\ref{fig:fig2a} 
for a neutrino energy of $E=150~\gev$.
}
\label{fig:fig2b}
\end{center}
\end{figure*}
}

%%%%%%%%%%%%%%%%%%%%%%%%%%%%%%%%%%%%%%%%%%%%%%%%%%%%%%%%%%%%%%%%%%%%%%%
%   `A2' FIT
%
\def\figAAc{
\begin{figure*}[!t]
\begin{center}
\includegraphics[width=1.80\columnwidth]{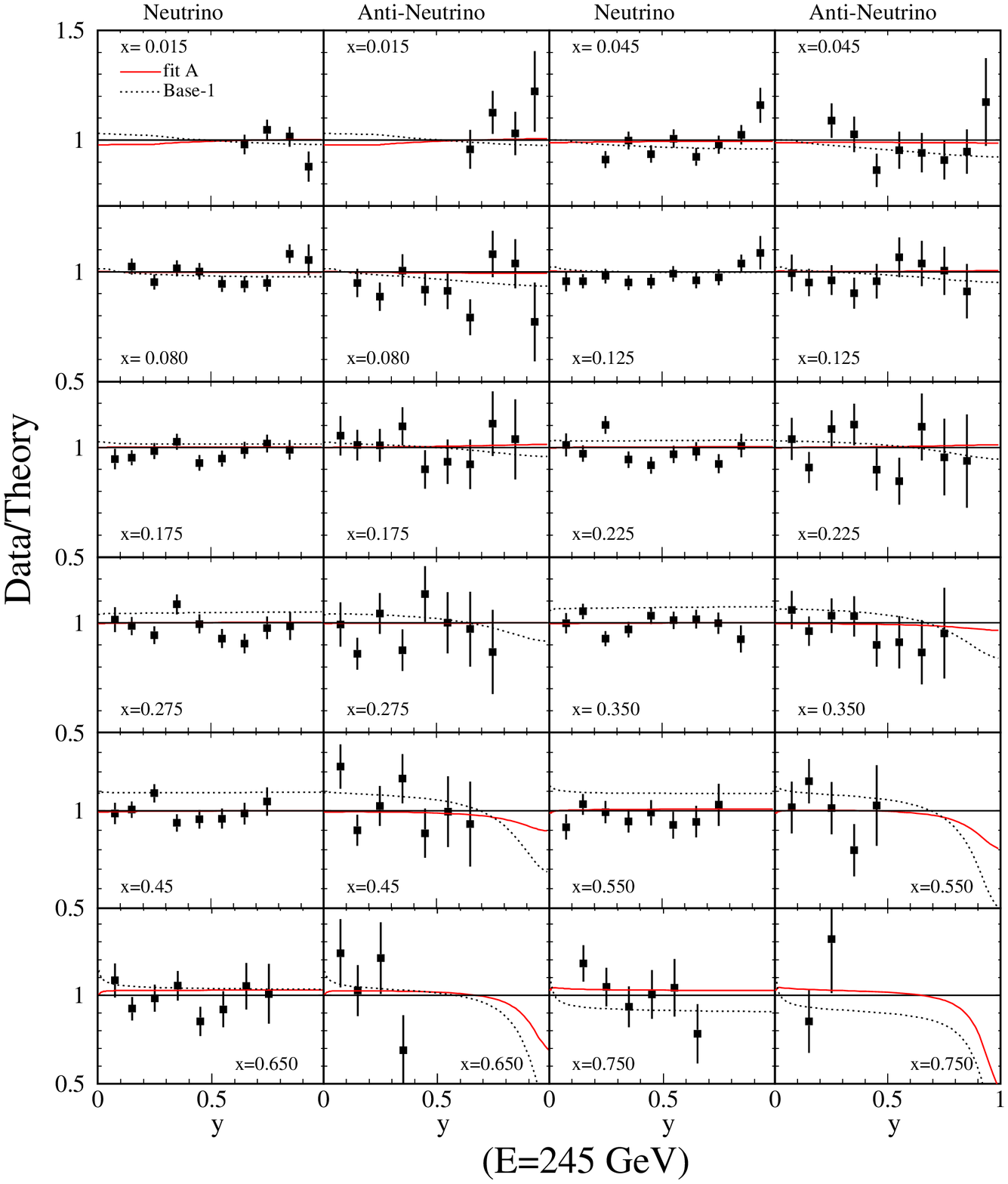} 
%\vspace*{-1cm}
\caption{
The same as in Fig.~\protect\ref{fig:fig2a} 
for a neutrino energy of $E=245~\gev$.
}
\label{fig:fig2c}
\end{center}
\end{figure*}
}

%%%%%%%%%%%%%%%%%%%%%%%%%%%%%%%%%%%%%%%%%%%%%%%%%%%%%%%%%%%%%%%%%%%%%%%
%   `A2' FIT PDF PLOTS
%
\def\figPDF{
\begin{figure}[!t]
\begin{center}
%
% Yellow band very intense
\includegraphics[width=0.95\columnwidth]{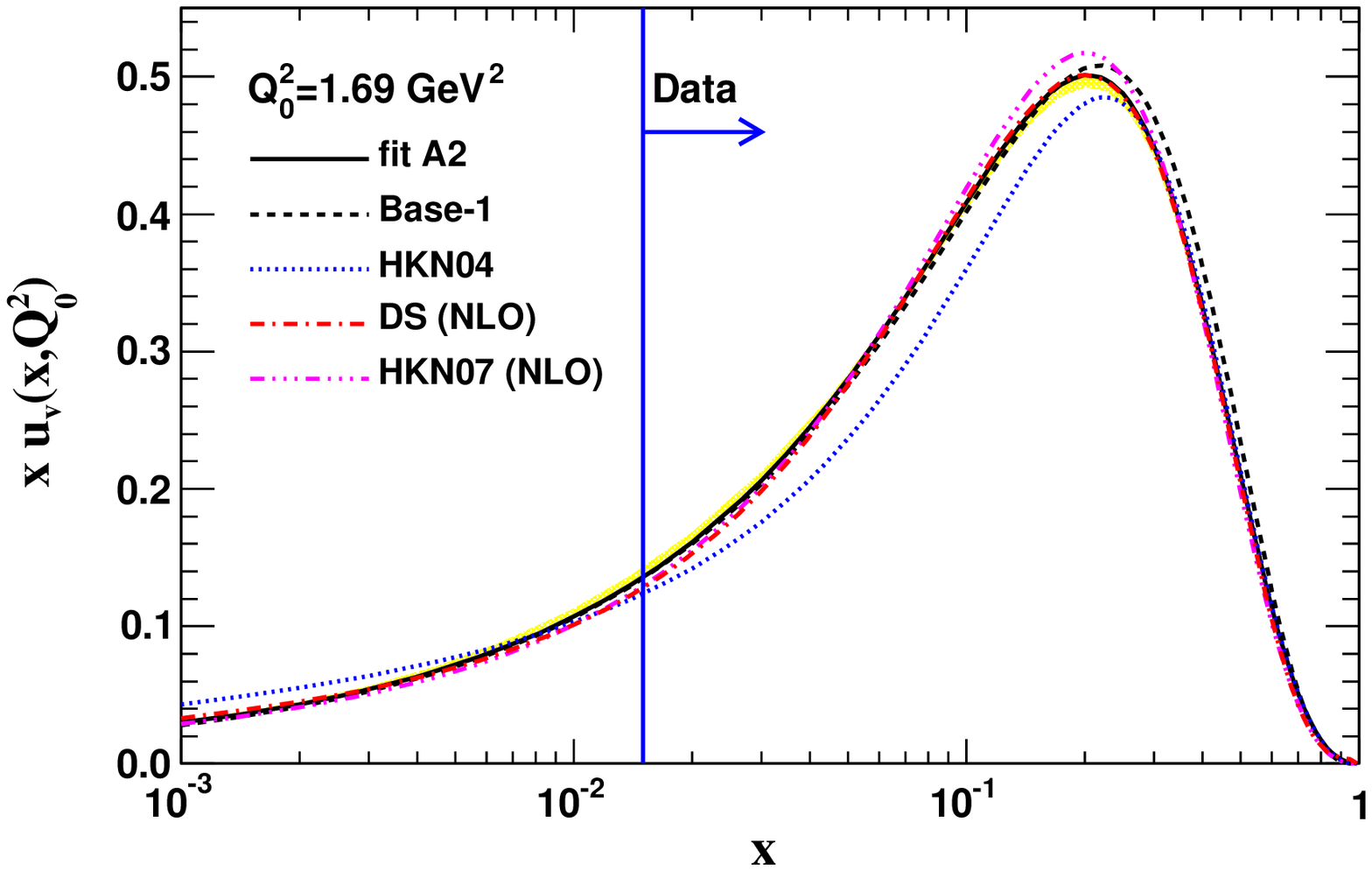} \\[-0.5cm]
\includegraphics[width=0.95\columnwidth]{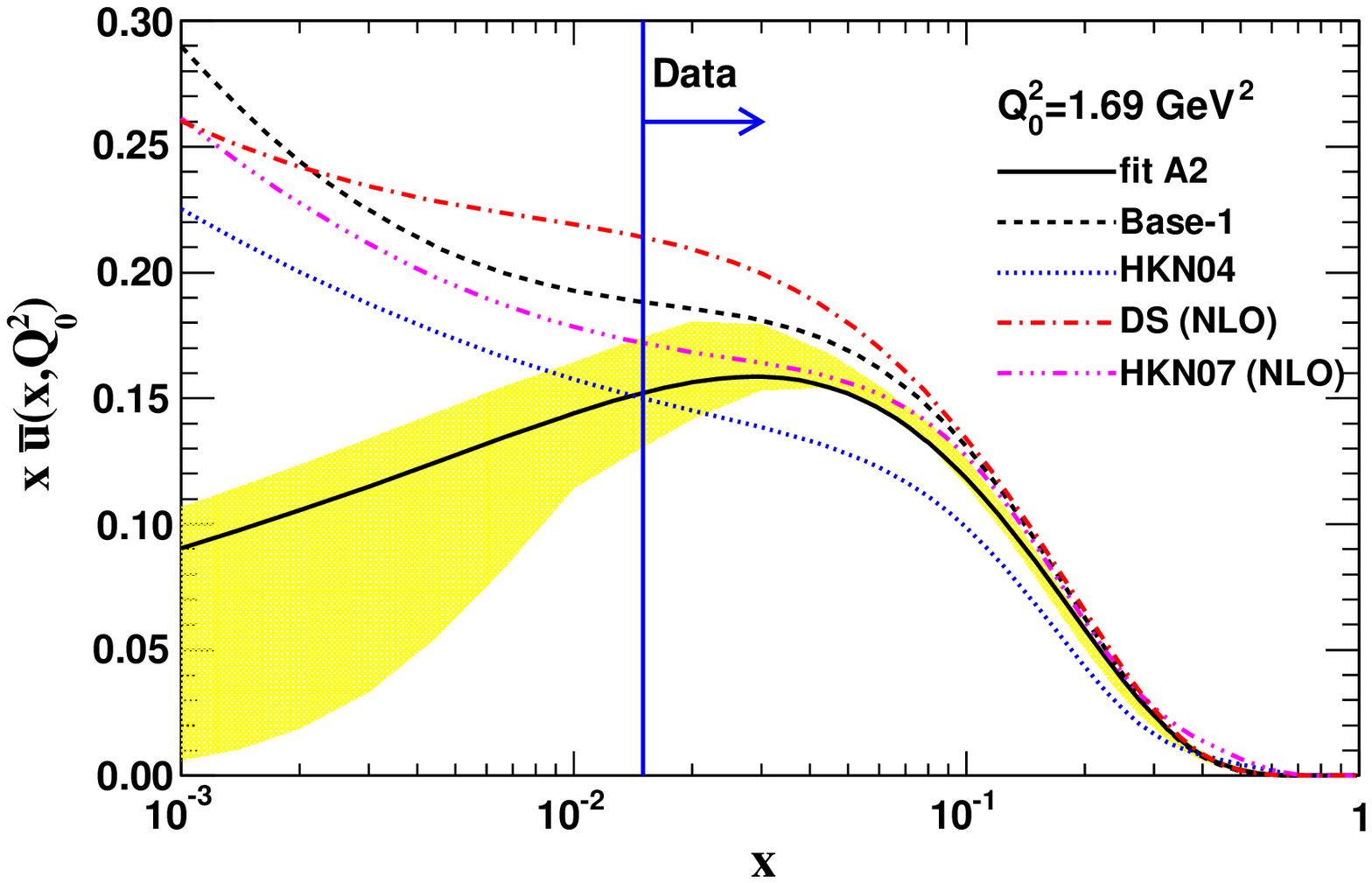} \\[-0.5cm]
\includegraphics[width=0.95\columnwidth]{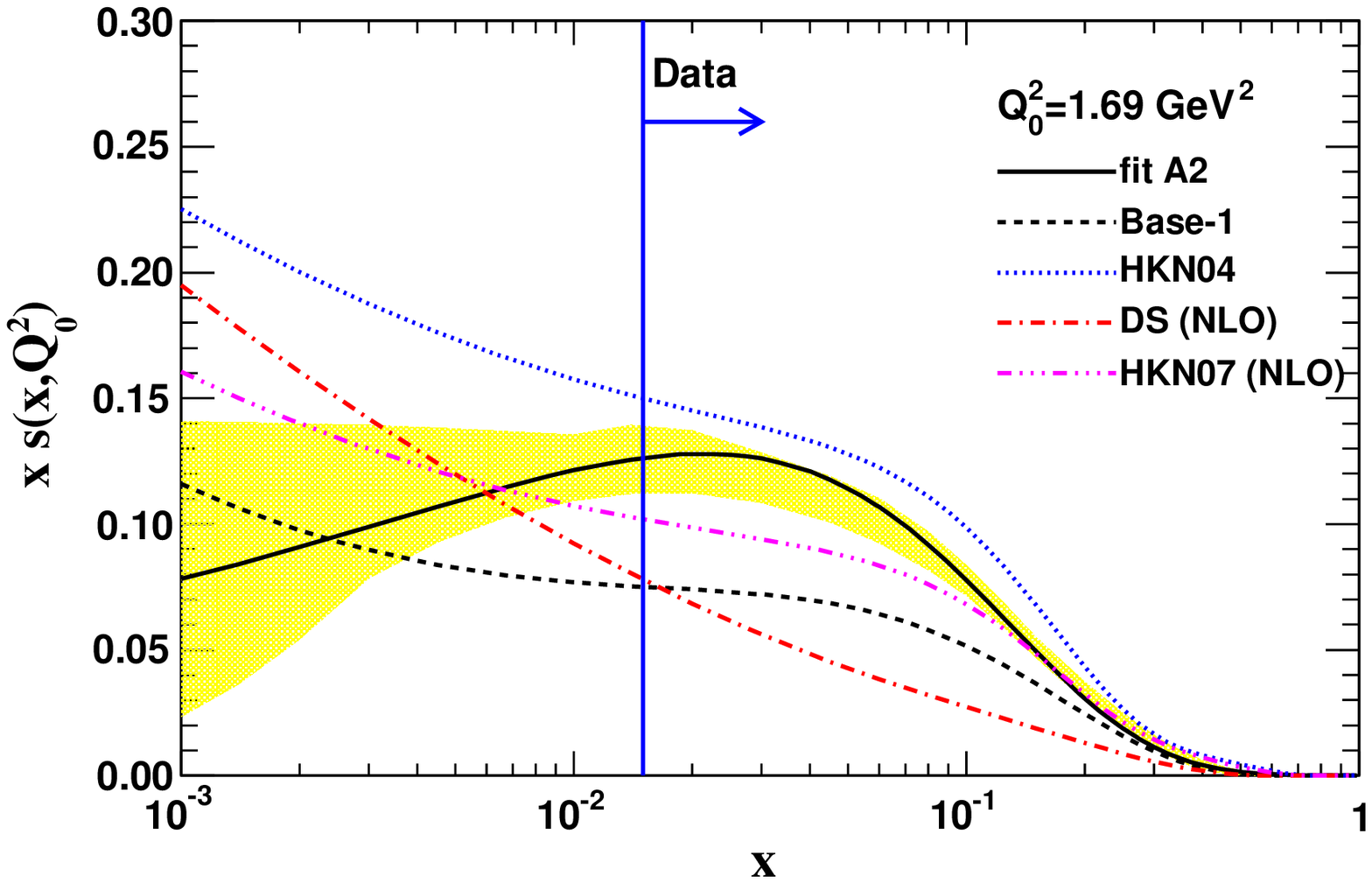} 
%
% Yellow band less intense
%\includegraphics[height=0.25\textheight]{eps/newfig4a2} 
%\includegraphics[height=0.25\textheight]{eps/newfig4b2} 
%\includegraphics[height=0.25\textheight]{eps/newfig4c2} 
\caption{
%\color{red}
Parton distributions for iron at our input scale $Q^2=1.69~\gevsq$.
Shown are the bands (in yellow) from fit `A2' for the up quark
valence distribution (upper figure), the up quark sea (middle), and
the strange quark sea (lower figure).  
The central PDF from fit `A2' is shown by the solid line.
The dashed lines depict parton
distributions constructed according to Eq.~\protect\eqref{eq:pdf} with
$A=56$ and $Z=26$ using the Base-1 
 free-proton PDFs.  The dotted lines
are the leading order HKN04 nuclear parton distributions
\protect\cite{Hirai:2004wq}, 
the dotted-dashed lines
are the next-to-leading order (NLO) HKN07 nuclear parton distributions
\protect\cite{Hirai:2007sx},
and the dot-dashed lines are the
next-to-leading order distributions (DS) from
Ref.~\protect\cite{deFlorian:2003qf}.  The vertical line marks the
lower limit of the data in the $x$ variable.
}
\label{fig:fig3b}
\end{center}
\end{figure}
}

%%%%%%%%%%%%%%%%%%%%%%%%%%%%%%%%%%%%%%%%%%%%%%%%%%%%%%%%%%%%%%%%%%%%%%%
%   D CORRECTION PLOT
%
\def\figDeuteron{
\begin{figure}[!t]
\begin{center}
\includegraphics[width=0.95\columnwidth]{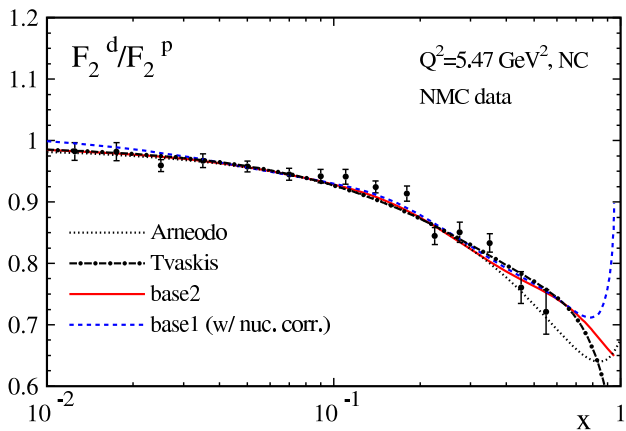} 
\caption{
NMC data for $F_2^D / F_2^p$ \protect\cite{Arneodo:1996kd}
at $Q^2 = 5.47~\gevsq$ in comparison with the theory prediction for
$F_2^D / F_2^p$ computed using free-proton Base-2 PDFs.
The dashed line shows the structure function ratio 
obtained with the Base-1 PDFs; in this case a nuclear correction
factor for deuterium has been applied 
({\it cf.},  Refs.~\protect\cite{Gomez:1993ri,Owens:2007kp}).
For comparison, we also show the parameterizations of 
Arneodo et al.\ \protect\cite{Arneodo:1996kd}
and 
Tvaskis et al.\ \protect\cite{Tvaskis:2004qm,Tvaskis:2006tv}.
}
%\caption{
%NMC data for $F_2^D / F_2^p$ \protect\cite{Arneodo:1996kd}
%at $Q^2 = 5.47~\gevsq$ in a data-over-theory representation
%where the 'theory' for $F_2^D / F_2^p$ has been computed 
%using free-proton Base-2 PDFs.
%The dashed line shows the structure function ratio 
%computed with the Base-1 PDFs; in this case a nuclear correction
%factor for deuterium has been applied 
%({\it cf.},  Refs.~\protect\cite{Gomez:1993ri,Owens:2007kp}).
%The resulting theory prediction has been normalized
%to the result obtained with the Base-2 PDFs.
%For comparison, we also show the parameterizations of 
%Arneodo et al.\ \protect\cite{Arneodo:1996kd}
%and 
%Tvaskis et al.\ \protect\cite{Tvaskis:2004qm,Tvaskis:2006tv}.
%}
\label{fig:fig4}
\end{center}
\end{figure}
}

%%%%%%%%%%%%%%%%%%%%%%%%%%%%%%%%%%%%%%%%%%%%%%%%%%%%%%%%%%%%%%%%%%%%%%%
%   SLAC/NMC PLOT
%
\def\figSLAC{
\begin{figure}[!t]
\begin{center}
\includegraphics[width=0.95\columnwidth]{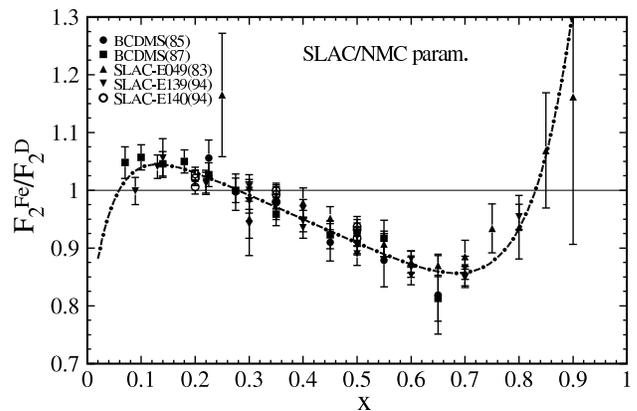} 
\caption{
Parameterization for the neutral current charged lepton structure function
$F_2^{Fe}/F_2^{D}$.
For comparison we show experimental results from the BCDMS
collaboration (BCDMS-85 \cite{Bari:1985ga}, BCDMS-87
\cite{Benvenuti:1987az}) and from experiments at SLAC (SLAC-E049
\cite{Bodek:1983qn}, SLAC-E139 \cite{Gomez:1993ri}, and SLAC-E140
\cite{Dasu:1993vk}).  Normalization uncertainties of the 
data have not been included.
}
\label{fig:fig5}
\end{center}
\end{figure}
}

%%%%%%%%%%%%%%%%%%%%%%%%%%%%%%%%%%%%%%%%%%%%%%%%%%%%%%%%%%%%%%%%%%%%%%%
%   R PLOT:  D SIGMA/DX DQ
%
\def\figDSIG{
\begin{figure}[!t]
\begin{center}
\includegraphics[width=0.95\columnwidth]{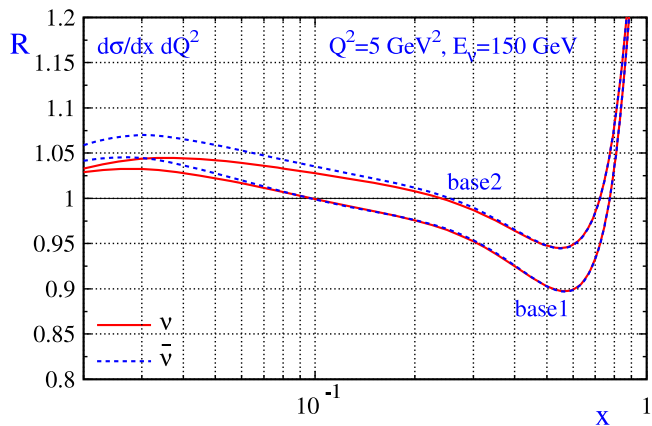} 
\caption{ 
Nuclear correction factor $R$ according to Eq.~\protect\eqref{eq:R}
for the differential cross section $d^2\sigma/dx \, dQ^2$ in charged
current $\nu Fe$ scattering at $Q^2=5~\gevsq$ and $E_\nu=150~\gev$.
 Results are shown 
using the `A2' fit 
for the charged current neutrino (solid lines)
and anti-neutrino (dashed lines) scattering from iron.
 The upper (lower) pair of curves shows the result of our analysis
with the Base-2 (Base-1) free-proton PDFs.
The correction factors shown here are for an iron target which has been corrected
for the neutron excess.
}
\label{fig:fig8}
\end{center}
\end{figure}
}

%%%%%%%%%%%%%%%%%%%%%%%%%%%%%%%%%%%%%%%%%%%%%%%%%%%%%%%%%%%%%%%%%%%%%%%
%   R PLOT:  NU
%
\def\figNU{
\begin{figure}[!t]
\begin{center}
\includegraphics[width=0.95\columnwidth]{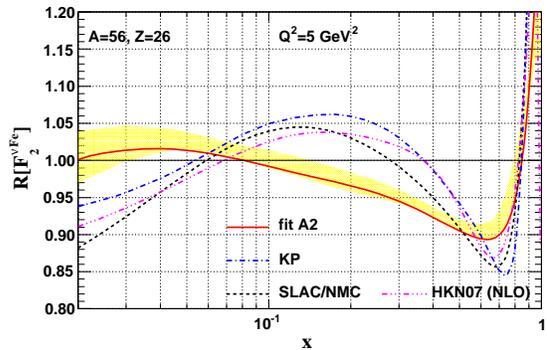} 
%\includegraphics[width=0.95\columnwidth]{eps/newfig8a_v3} 
%\\[15pt]
\\[20pt]
\includegraphics[width=0.95\columnwidth]{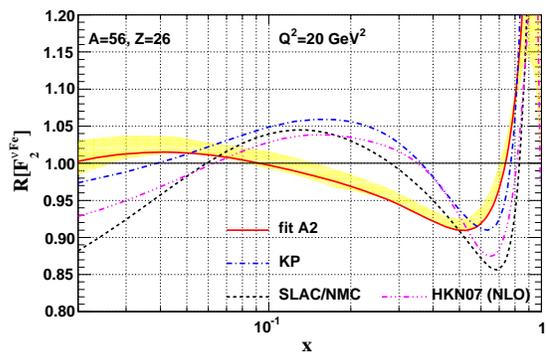} 
\\[15pt]
\caption{ 
Nuclear correction factor $R$ according to Eq.~\protect\eqref{eq:R} for the
structure function $F_2$ in charged current $\nu Fe$ scattering at 
a)~$Q^2=5~\gevsq$ and b)~$Q^2=20~\gevsq$.  The  solid curve
shows the result of our analysis of NuTeV data (one representative of
'fit A2') divided by the results obtained with the Base-1 
free-proton PDFs; 
the uncertainty from the A2 fit is represented by the yellow band. 
For comparison we show the correction factor from
the Kulagin--Petti model 
\protect\cite{Kulagin:2004ie} (dashed-dot line), 
HKN07  \protect\cite{Hirai:2007sx} (dashed-dotted line), 
and the SLAC/NMC parametrization (dashed line)  
\protect\cite{NMC}.
%We compute this for $\{A=56,Z=26\}$.
%
}
\label{fig:fig6a}
\end{center}
\end{figure}
}

%%%%%%%%%%%%%%%%%%%%%%%%%%%%%%%%%%%%%%%%%%%%%%%%%%%%%%%%%%%%%%%%%%%%%%%
%   R PLOT:  NU-BAR
%
\def\figNUBAR{
\begin{figure}[!t]
\begin{center}
\includegraphics[width=0.95\columnwidth]{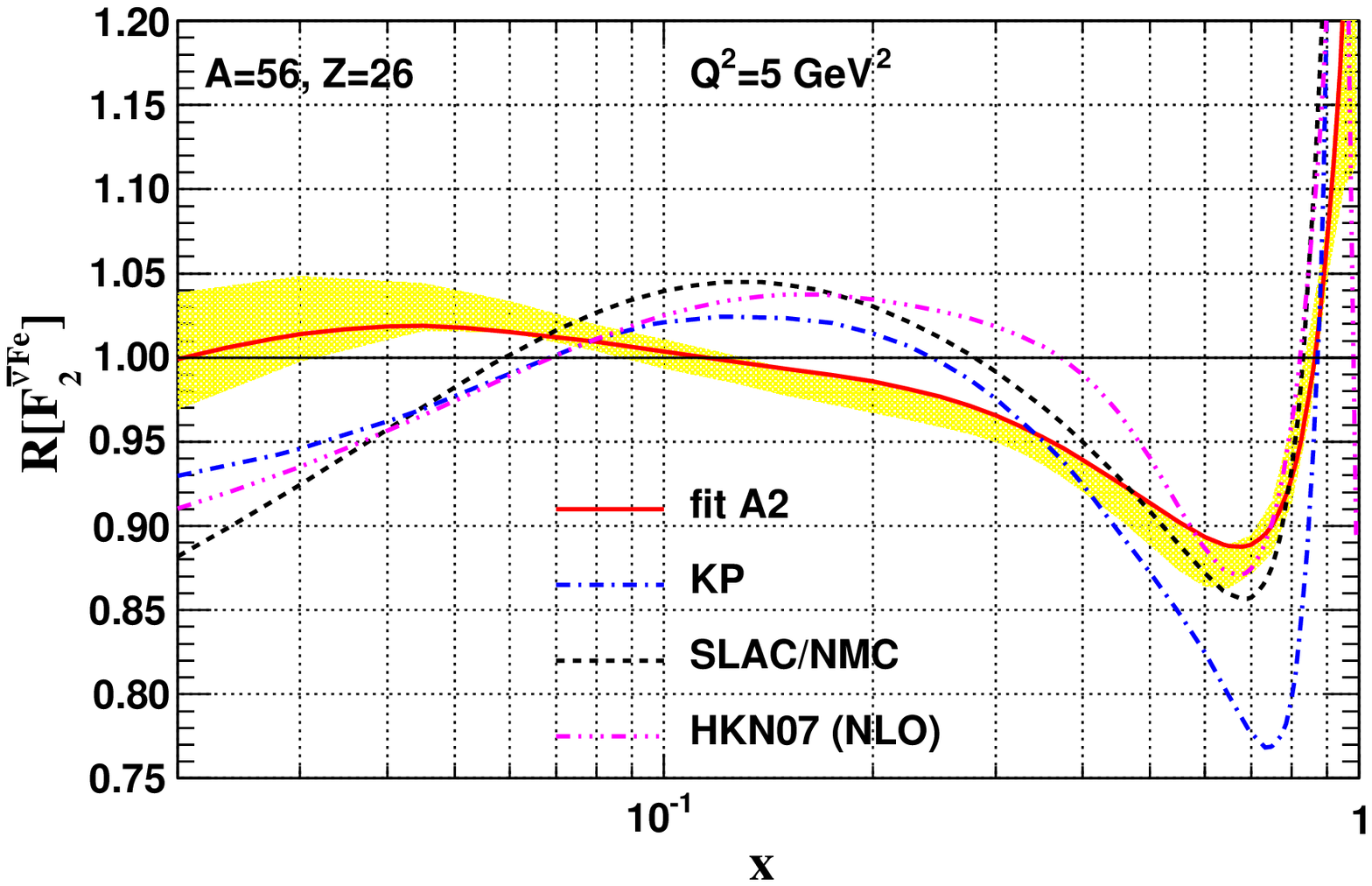} 
%\includegraphics[width=0.95\columnwidth]{eps/newfig9a_v3} 
%\\[15pt]
\\[20pt]
\includegraphics[width=0.95\columnwidth]{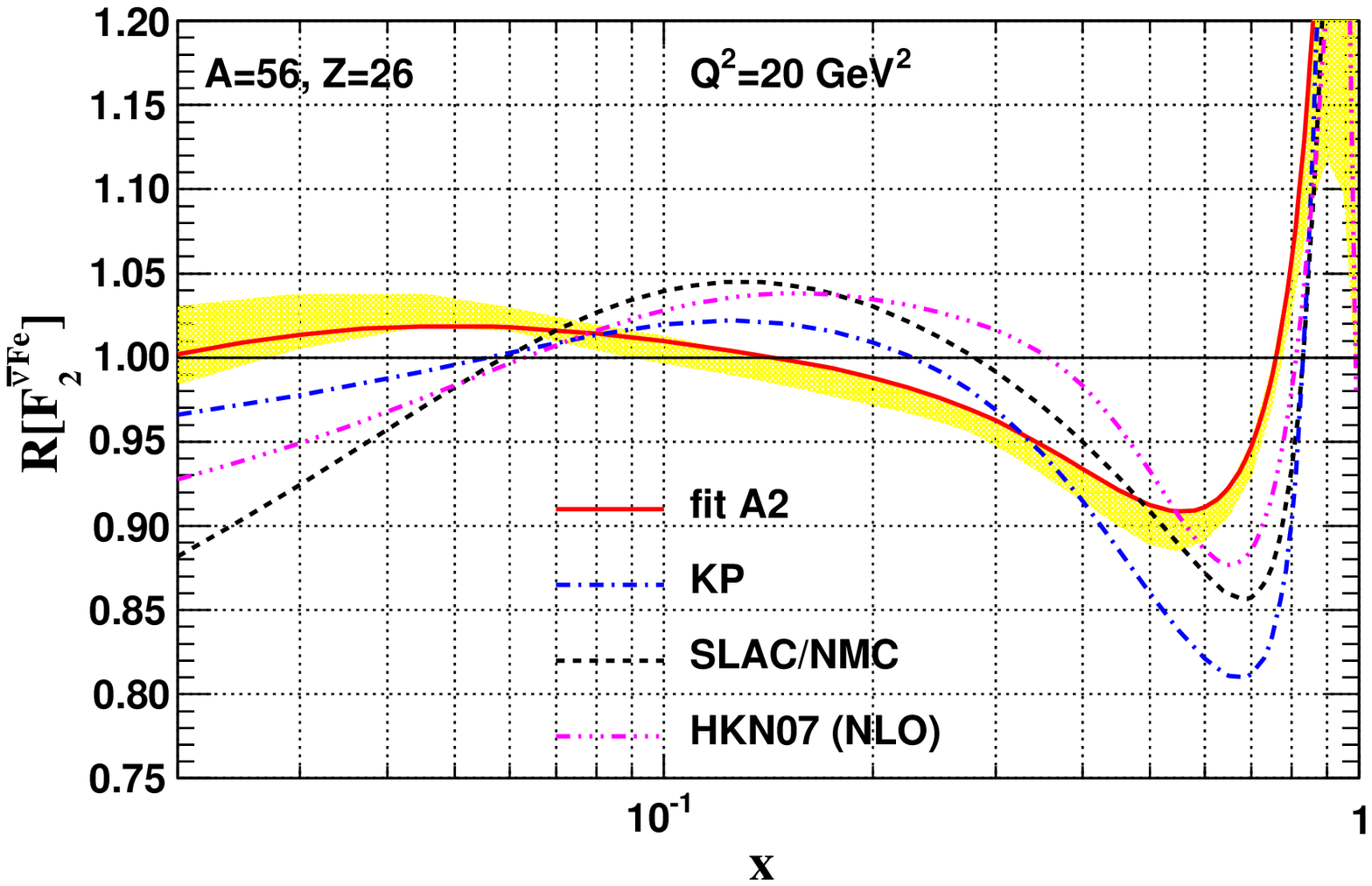} 
\\[15pt]
\caption{
The same as in Fig.~\protect\ref{fig:fig6a} for $\bar{\nu} Fe$
scattering.
%We compute this for $\{A=56,Z=26\}$.
}
\label{fig:fig6b}
\end{center}
\end{figure}
}

%%%%%%%%%%%%%%%%%%%%%%%%%%%%%%%%%%%%%%%%%%%%%%%%%%%%%%%%%%%%%%%%%%%%%%%
%   SLAC/NMC PARM VS OUR CURVES 
%
\def\figFINAL{
\begin{figure}[!t]
\begin{center}
\includegraphics[width=0.95\columnwidth]{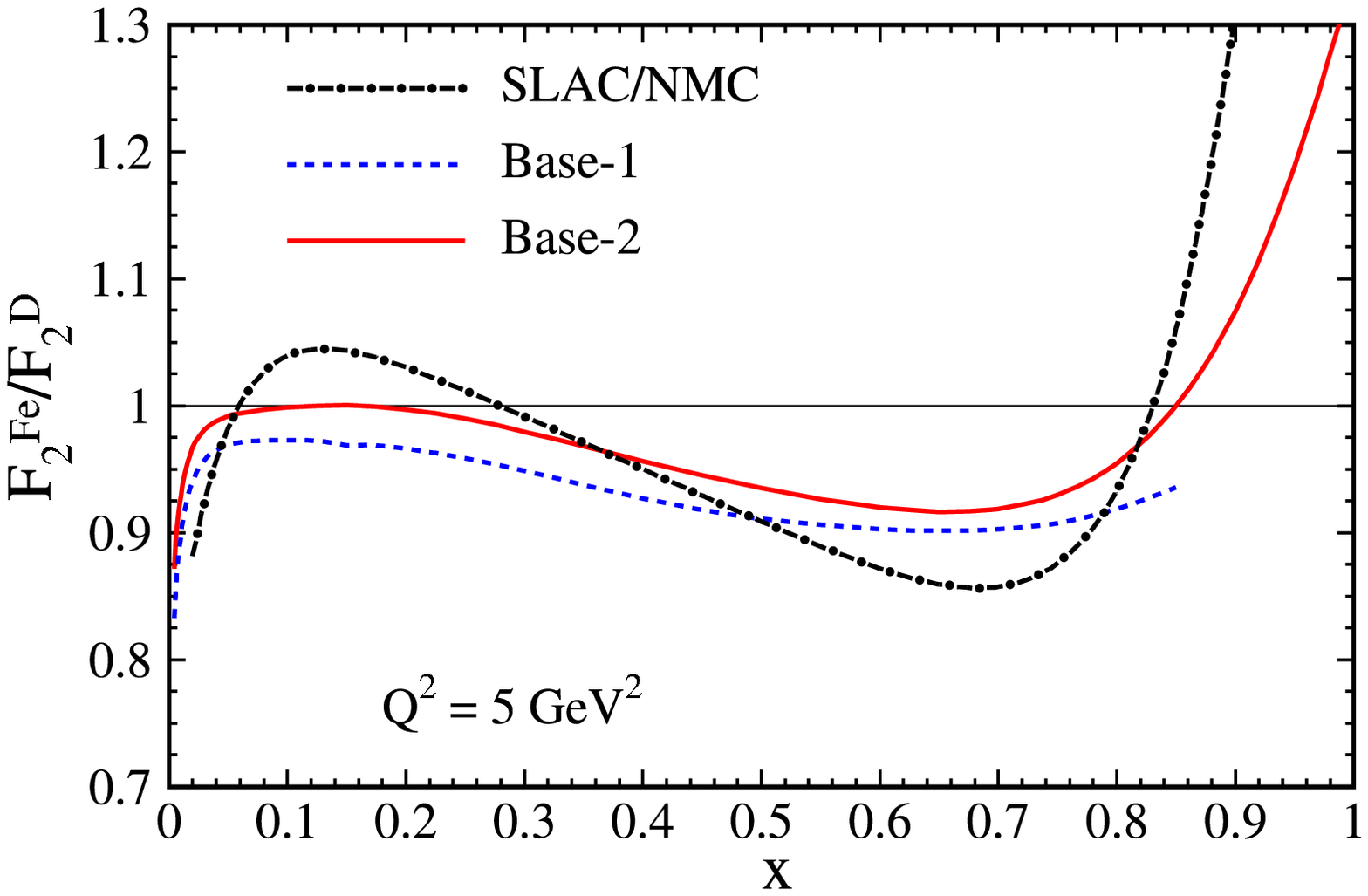} 
\caption{ 
Predictions (solid and dashed line) for the  structure function ratio 
$F_2^{Fe}/F_2^{D}$ using the iron PDFs extracted from fits to NuTeV
neutrino and anti-neutrino data (fit `A2').
The SLAC/NMC parameterization is shown with the dot-dashed line. 
The structure function $F_2^D$ in the denominator has been computed
using either the Base-2 (solid line) or the Base-1 (dashed line) PDFs.
A nuclear correction factor for deuterium
 has been included in the Base-1 calculation \protect\cite{Owens:2007kp}.
}
\label{fig:ccncCompare}
\end{center}
\end{figure}
}

%%%%%%%%%%%%%%%%%%%%%%%%%%%%%%%%%%%%%%%%%%%%%%%%%%%%%%%%%%%%%%%%%%%%%%%

%**********************************************************
\clearpage
\tableofcontents

%**********************************************************
\clearpage
\section{Introduction}
\label{sec:intro}

The high statistics measurements of neutrino
deeply inelastic scattering (DIS) on heavy nuclear targets 
has generated significant interest  in the
literature since these measurements provide valuable information for global fits
of parton distribution functions (PDFs) \cite{Thorne:2006wq}.
The use of nuclear targets is unavoidable due to the weak nature of the
neutrino interactions, and this complicates the extraction of free nucleon
PDFs because model-dependent corrections must be applied to the data.

Additionally, these  same data are also useful for extracting the 
{\em nuclear} parton distribution functions (NPDFs); 
for such an analysis,  no nuclear correction factors are required.
Due to the limited statistics available for individual nuclear targets
with a given atomic number $A$ the standard approach is to model the
$A$-dependence of the fit parameters,  and then combine the data sets for
many different target materials in the global 
analysis \cite{Hirai:2001np,Hirai:2004wq,Hirai:2007sx,Eskola:1998df,Eskola:2007my,deFlorian:2003qf}.
However, the high statistics NuTeV neutrino--iron cross section data
($> 2000$ points) offer the possibility to investigate the viability
of a dedicated determination of iron PDFs \cite{Tzanov:2005kr}.

With this motivation, we will perform a fit to the NuTeV
neutrino--iron data and extract the corresponding iron PDFs.
Since we are studying iron alone and will not (at present) combine the
data with measurements on different target materials, we need not make
any assumptions about the nuclear corrections; this side-steps a
number of difficulties \cite{Gomez:1993ri,Owens:2007kp,Thorne:2006zu}.

While this approach has the advantage that we do not need to model the
$A$-dependence, it has the drawback that the data from just one
experiment will not be sufficient to constrain all the parton
distributions. Therefore, other assumptions must enter the
analysis. The theoretical framework will roughly follow the CTEQ6
analysis of free proton PDFs \cite{Pumplin:2002vw}; this will be
discussed in Sec.~\ref{sec:framework}.

In Sec.~\ref{sec:iron} we present the results of our analysis, and
compare with nuclear PDFs from the literature.
In Sec.~\ref{sec:nfac} we extract the nuclear correction factors from
our iron PDFs and compare with a SLAC/NMC parameterization taken from
the $\ell^\pm$--Fe DIS process \cite{NMC} 
and also with the
parameterization by Kulagin \&
Petti \cite{Kulagin:2004ie,Kulagin:2007ju}.
Finally, we summarize our results and conclusions in
Sec.~\ref{sec:summary}.

%\clearpage
%%%%%%%%%%%%%%%%%%%%%%%%%%%%%%%%%%%%%%%%%%%%%%%%%%%%%%%%%%%%%%%%%%%%
\section{Theoretical Framework}
\label{sec:framework}
\subsection{Basic formalism}

For our PDF analysis, we will use the  general features of the 
QCD-improved parton model
and the $\chi^2$ analyses as outlined in Ref.~\cite{Pumplin:2002vw}.
Here, we will focus on  
the issues specific to our 
study of NuTeV neutrino--iron data in terms
of nuclear parton distribution functions.
We adopt the framework of the recent CTEQ6 analysis of proton PDFs 
where the input distributions at the scale $Q_0=1.3~\gev$ 
are parameterized as \cite{Pumplin:2002vw}
%\clearpage
\begin{widetext}
\begin{eqnarray}
x f_i(x,Q_0) =
\begin{cases}
A_0 x^{A_1} (1-x)^{A_2} e^{A_3 x}(1+ e^{A_4} x)^{A_5} 
&: i=u_v,d_v,g,\bar{u}+\bar{d},s,\bar{s} \, ,
\\
A_0 x^{A_1} (1-x)^{A_2} + (1 + A_3 x) (1- x)^{A_4} 
&: i = \bar{d}/\bar{u}  \, ,
\end{cases}
\label{eq:input}
\end{eqnarray} 
\end{widetext}
where $u_v$ and $d_v$ are 
the up- and down-quark valence distributions,
$\bar{u}$, $\bar{d}$, $s$, $\bar{s}$ are the up, 
down, strange and anti-strange sea distributions,
and $g$ is the gluon.
Furthermore, the $f_i=f_i^{p/A}$ denote parton distributions of {\em
bound protons in the nucleus $A$}, and the variable $0 \le x \le A$ is
defined as $x := A x_A$ where $x_A = Q^2/2 p_A \cdot q$ is the usual
Bjorken variable formed out of the four-momenta of the nucleus and the
exchanged boson.
Equation \eqref{eq:input} is designed for $0 \le x \le 1$ and we here
neglect\footnote{%
While the nuclear PDFs can be finite for $x>1$, the
magnitude of the PDFs in this region is negligible for
the purposes of the present study ({\it cf.},
Refs.~\protect\cite{Hirai:2001np,Hirai:2004wq,Hirai:2007sx,Eskola:1998df,Eskola:2007my}).
}
the distributions at $x>1$.
Note that the condition $f_i(x>1,Q) = 0$ 
is preserved by the DGLAP evolution and
has the effect that the evolution equations and sum rules
for the $f_i^{p/A}$
are the same as in the free proton case.\footnote{%
While the quark number and momentum sum rules for the nuclear case are satisfied as in the proton, 
there is no requirement that the 
momentum fractions carried by the PDF flavors be the same. 
%}%%% END FOOTNOTE
%
%\footnote{%
A recent analysis at low $Q^2$
found the Cornwall-Norton moments to be the same in
iron as in deuterium formed from a free proton and a free
neutron to within $3\%$ \protect\cite{Niculescu:2005rh}.
}%%% END FOOTNOTE

The PDFs for a nucleus $(A,Z)$ are constructed as
\begin{equation}
f_i^A(x,Q) = \frac{Z}{A}\  f_i^{p/A}(x,Q) + \frac{(A-Z)}{A}\  f_i^{n/A}(x,Q)
\label{eq:pdf}
\end{equation}
where we relate the distributions inside a bound neutron, $f_i^{n/A}(x,Q)$, to the
ones in a proton by assuming isospin symmetry. 
Similarly, the nuclear structure functions are given by
\begin{equation}
F_i^A(x,Q) =  \frac{Z}{A}\  F_i^{p/A}(x,Q) + \frac{(A-Z)}{A}\  F_i^{n/A}(x,Q)
\label{eq:sfs}
\end{equation}
such that they can be computed in next-to-leading order 
as convolutions of the nuclear PDFs with the conventional
Wilson coefficients, {\it i.e.}, generically 
\begin{equation}
F_i^A(x,Q) = \sum_k C_{ik} \otimes f_k^{A}\, .
\label{eq:sfs2}
\end{equation}
In order to take into account heavy quark mass effects
we calculate the relevant structure functions
in the ACOT scheme \cite{Aivazis:1993kh,Aivazis:1993pi} in NLO QCD
\cite{Kretzer:1998ju}. 
Finally, the differential cross section for charged current 
(anti-)neutrino--nucleus scattering is given in terms of three
structure functions:
\begin{eqnarray}
\frac{d^2 \sigma}{dx\, dy}^{\overset{(-)}{\nu}A}
&=& \frac{G^2 M E}{\pi} \left[ 
(1-y-\frac{M x y}{2 E}) F_2^{\overset{(-)}{\nu}A} 
\right.
\nonumber \\
&+& 
\left.
\frac{y^2}{2} 2 x F_1^{\overset{(-)}{\nu}A} 
\pm y(1-\frac{y}{2}) x F_3^{\overset{(-)}{\nu}A}
\right]\, ,
\label{eq:dsig}
\end{eqnarray}
where the '$+$' ('$-$') sign refers to neutrino (anti-neutrino) scattering
and where $G$ is the Fermi constant, $M$ the nucleon mass, and $E$ the energy of the
incoming lepton (in the laboratory frame).

%%%%%%%%%%%%%%%%%%%%%%%%%%%%%%%%%%%%%%%%%%%%%%%%%%%%%%%%%%%%%%%%%%%%
%%%%%%%%%%%%%%%%%%%%%%%%%%%%%%%%%%%%%%%%%%%%%%%%%%%%%%%%%%%%%%%%%%%%

\subsection{Constraints on PDFs}

We  briefly discuss which combinations of
PDFs can be constrained by the neutrino--iron data.
For simplicity, we restrict ourselves to leading order, neglect
heavy quark mass effects (as well as the associated production thresholds),
and assume a diagonal Cabibbo-Kobayashi-Maskawa (CKM) matrix.\footnote{All 
these effects are properly included in our calculations.}
 The neutrino--iron structure
functions are given by (suppressing the dependence on $x$
  and $Q^2$):
\begin{eqnarray}
F_1^{\nu A} &=& d^A + s^A + \bar{u}^A + \bar{c}^A + \ldots \, ,
\\
F_2^{\nu A} &=& 2 x F_1^{\nu A}\, ,
\label{eq:f2nu}
\\
F_3^{\nu A} &=& 2\left [d^A + s^A - \bar{u}^A - \bar{c}^A + \ldots \right] \, .
\end{eqnarray}
The structure functions for anti-neutrino scattering are obtained by exchanging
the quark and anti-quark PDFs in the corresponding neutrino structure functions:
\begin{eqnarray}
F_{1,2}^{\bar{\nu}A} &=& +F_{1,2}^{\nu A}[q \leftrightarrow \bar{q}]\  , \,
  \\
F_{3}^{\bar{\nu}A} &=& -F_{3}^{\nu A}[q \leftrightarrow \bar{q}] \  .
\end{eqnarray}
Explicitly this gives
\begin{eqnarray}
F_1^{\bar\nu A} &=& u^A + c^A + \bar{d}^A + \bar{s}^A + \ldots \, ,
\\
F_2^{\bar\nu A} &=& 2 x F_1^{\bar\nu A}\, ,
\label{eq:f2nub}
\\
F_3^{\bar\nu A} &=& 2\left [u^A + c^A - \bar{d}^A - \bar{s}^A + \ldots \right] \, .
\end{eqnarray}

It is instructive to compare this with the parton model expressions for
the structure function $F_{2}$ in $l^\pm A$ scattering, where $l^\pm$ denotes
a charged lepton:
\begin{eqnarray}
\frac{1}{x}\ F_2^{lA}  &=& 
\frac{4}{9} (u^A + \bar{u}^A) + 
\frac{1}{9} (d^A + \bar{d}^A) + 
\nonumber \\ &+&
\frac{1}{9} (s^A + \bar{s}^A) + 
\frac{4}{9} (c^A + \bar{c}^A) + \ldots \, .
\label{eq:f2em}
\end{eqnarray}
Using the Callan--Gross relations in Eqs.~\eqref{eq:f2nu} and \eqref{eq:f2nub},
and neglecting the proton mass, the differential cross section Eq.~\eqref{eq:dsig}
can be  simplified in the form
\begin{eqnarray}
\der \sigma &\propto& (1-y+y^2/2) F_2^{\overset{(-)}{\nu}A} 
\pm y(1-\frac{y}{2}) x F_3^{\overset{(-)}{\nu}A} 
\end{eqnarray}
with the limiting cases: 
\begin{eqnarray}
\der \sigma
&\longrightarrow& 
\begin{cases}
\frac{1}{2} F_2^{\overset{(-)}{\nu}A} \pm \frac{1}{2} x F_3^{\overset{(-)}{\nu}A}
& ({\rm for}\,\, y \to 1)
\\[10pt]
 F_2^{\overset{(-)}{\nu}A}
& ({\rm for}\,\, y \to 0)
\end{cases}
\end{eqnarray}
The latter form of $\der \sigma$ shows that the (anti-)neutrino cross section data
naturally encodes information on the four structure function combinations
$F_2^{\overset{(-)}{\nu}A} \pm x F_3^{\overset{(-)}{\nu}A}$ and 
$F_2^{\overset{(-)}{\nu}A}$ in separate regions of the phase space.

If we assume\footnote{Note that these equations are known {\it not} to be
exact as the DGLAP evolution equations at NNLO generate an asymmetry
even if one starts with $s=\bar s$ or $c=\bar c$ at some scale
$Q^2$ \cite{Catani:2004nc}. However, these effects are tiny and far
beyond the accuracy of our study.}  
$s^A=\bar{s}^A$ and $c^A=\bar{c}^A$,
the structure functions $F_2^{\overset{(-)}{\nu}A}$ constrain the valence distributions
$d_v^A = d^A - \bar{d}^A$, $u_v^A = u^A - \bar{u}^A$ and the flavor-symmetric sea
$\Sigma^A := \bar{u}^A + \bar{d}^A + \bar{s}^A + \bar{c}^A + \ldots$ via the relations:
\begin{eqnarray}
\frac{1}{x} F_2^{\nu A} &=& 2 \left[d_v^A + \Sigma^A \right]\, , \qquad
\\
\frac{1}{x} F_2^{\bar\nu A} &=& 2 \left[u_v^A + \Sigma^A \right]\, .
\end{eqnarray}
Furthermore, we have
\begin{eqnarray}
\frac{1}{x} F_2^{\nu A}+ F_3^{\nu A} &=& 4 (d^A + s^A)\, , \qquad
\\
\frac{1}{x} F_2^{\bar\nu A}- F_3^{\bar\nu A}&=& 4 (\bar{d}^A + \bar{s}^A)\, .
\end{eqnarray}
Since we constrain the strange distribution utilizing the dimuon data,\footnote{See
Refs.~\cite{Goncharov:2001qe,Mason:2005vc,Mason:2004yf,Kretzer:2003wy,Olness:2003wz,Kretzer:2001tc}
for details.} the latter two structure functions are useful to separately extract the 
$d^A$ and $\bar{d}^A$ distributions.

For an isoscalar nucleus we encounter further simplifications. In this case,
$u^A=d^A$ and $\bar{u}^A = \bar{d}^A =:\bar{q}^A$ which implies $u_v^A = d_v^A =: v^A$.
Hence, the independent quark distributions are 
$\{ v^A, \bar{q}^A, s^A=\bar{s}^A,  c^A=\bar{c}^A,\,  \ldots \}$.
It is instructive to introduce the parameter $\Delta := 1/2-Z/A$ which describes the
degree of non-isoscalarity.
This allows us to write the PDFs in a way which makes deviations from 
isoscalarity manifest:
\begin{eqnarray}
u_v^A &=& v^A -\Delta [u_v^{p/A} - d_v^{p/A}]
\label{eq:deviation1}
\\
d_v^A &=& v^A + \Delta [u_v^{p/A} - d_v^{p/A}]
\\
\bar{u}^A &=& \bar{q}^A - \Delta [\bar{u}^{p/A} - \bar{d}^{p/A}]
\\
\bar{d}^A &=& \bar{q}^A + \Delta [\bar{u}^{p/A} - \bar{d}^{p/A}]
\label{eq:deviation4}
\end{eqnarray}
in terms of an averaged nuclear valence distribution $v^A =
(u_v^{p/A}+d_v^{p/A})/2$ and an averaged nuclear sea distribution
$\bar{q}^A = (\bar{u}^{p/A}+\bar{d}^{p/A})/2$.  Recall, $f_i^{p/A}$
represents the distribution for a bound proton in the nucleus $A$;
hence, the nuclear effects are encoded in these terms.
Notice that non-isoscalar targets ($\Delta\not=0$)
therefore provide information on the
difference between the valence distributions ($u_v^{p/A} - d_v^{p/A}$)
and the light quark sea distribution ($\bar{u}^{p/A} - \bar{d}^{p/A}$)
in the nucleon.  Unfortunately, the data are often corrected for
non-isoscalar effects and this information is lost.

%%%%%%%%%%%%%%%%%%%%%%%%%%%%%%%%%%%%%%%%%%%%%%%%%%%%%%%%%%%%%%%%%%%%
%%%%%%%%%%%%%%%%%%%%%%%%%%%%%%%%%%%%%%%%%%%%%%%%%%%%%%%%%%%%%%%%%%%%

%%%%%%%%%%%%%%%%%%%%%%%%%%%%%%%%%%%%%%%%%%%%%%%%%%%%%%%%%%%%%%%%%%%%
\subsection{Methodology}
\label{method}

The basic formalism described in the previous sections is implemented
in a global PDF fitting package, {\it but} with the difference that no
nuclear corrections are applied to the analyzed data; hence, the
resulting PDFs are for a bound proton in an iron nucleus.
The parameterization of Eq.~\eqref{eq:input} provides enough
flexibility to describe current data sets entering a global analysis
of free nucleon PDFs; given that the nuclear modifications of the
$x$-shape appearing in this analysis are modest, this parameterization
will also accommodate the iron PDFs.

Because the neutrino data alone do not have the power to constrain all
of the PDF components, we will need to impose some minimal set of
external constraints.  For example, our results are rather insensitive
to the details of the gluon distribution with respect to both the
overall $\chi^2$ and also the effect on the quark distributions.
The nuclear gluon distribution is very weakly constrained by present
data, and a gluon PDF with small nuclear modifications has been found
in the NLO analysis of Ref.~\cite{deFlorian:2003qf}.
We have therefore fixed the gluon input parameters to their free
nucleon values.  For the same reasons the gluon is not sensitive to
this analysis, fixing the gluon will have minimal effect on our
results.
Furthermore, we have set the $\bar{d}/\bar{u}$ ratio to the free
nucleon result assuming that the nuclear modifications to the down and
up sea are similar such that they cancel in the ratio.
This assumption is supported by Fig.~6 in
Ref.~\cite{deFlorian:2003qf}.

Because we have limited the data set to a single heavy target (iron),
the $\chi^2$ surface has some parameter directions which are
relatively flat.  To fully characterize the parameter space, we
perform many ``sample fits'' starting from different initial
conditions, and iterate these fits including/excluding additional
parameters.  The result is a set of bands for fits of comparable
quality ($\Delta\chi^2\sim 50$ for 2691 data points) which provide an
approximate measure of the constraining power of the data.

%%%%%%%%%%%%%%%%%%%%%%%%%%%%%%%%%%%%%%%%%%%%%%%%%%%%%%%%%%%%%%%%%
\section{Analysis of iron data}
\label{sec:iron}

%%%%%%%%%%%%%%%%%%%%%%%%%%%%%%%%%%%%%%%%%%%%%%%%%%%%%%%%%%%%%%%%%%%%%%%%%%%%%5
\subsection{Iron Data Sets}

We determine iron PDFs using the recent NuTeV differential neutrino 
(1371/1170 data points) and 
anti-neutrino (1146/966 data points) DIS cross section
data \cite{Tzanov:2005kr} 
where
the quoted numbers of data points refer to the two different combinations of 
kinematic cuts introduced below.
In addition, we include NuTeV/CCFR dimuon data (174 points) \cite{Goncharov:2001qe} 
which are sensitive to the strange quark content of the nucleon.

There are other measurements of neutrino--iron DIS 
available in the literature 
from the CCFR \cite{Oltman:1992pq,Seligman:1997mc,Yang:2000ju,Yang:2001rm},
CDHS \cite{Abramowicz:1984yk} and CDHSW \cite{Berge:1989hr} collaborations; see, e.g., 
Ref.\ \cite{Conrad:1997ne} for a review.
There is also a wealth of charged lepton--iron DIS data including
SLAC \cite{Dasu:1993vk} and 
EMC \cite{Aubert:1986yn,Aubert:1987da}.\footnote{{\it Cf.} 
the Durham HEP Databases for a complete listing: {\tt http://www-spires.dur.ac.uk/hepdata/}
}  %%% END FOOTNOTE   
For the present study we limit our analysis to the NuTeV experiment alone; we will 
compare and contrast different experiments in  a later study. 
%

%%%%%%%%%%%%%%%%%%%%%%%%%%%%%%%%%%%%%%%%%%%%%%%%%%%%%%%%%%%%%%%%%%%%%%%%%%%%%5
\subsection{Fit results}
% Description of fits; table 1: Fits to NuTeV data

%%%%%%%%%%%%%%%%%%%%%%%%%%%%%%%%%%%%%%%%%%%%%%%%%%%%%%%%%%%%%%%%%%%%%%%%%%%%%%%%%%%%%%%
\begin{table}[t]
%\small
%\scriptsize
\begin{center}
\begin{tabular}{l|l|l|c|c|c|l}
\hline 
Scheme & Cuts       & Data  & $\#$ points       &$\chi^2$ & $\chi^2$/pts & Name \\
\hline
ACOT & $Q >1.3$ GeV  & $\nu+\bar{\nu}$  & 2691 & 3678 & 1.37 & A\\
     & no $W_{cut}$       & $\nu$               & 1459 & 2139 & 1.47 & A$\nu$\\
     &                    & $\bar{\nu}$         & 1232 & 1430 & 1.16 & A$\bar{\nu}$\\
\hline
ACOT & $Q >2$ GeV    & $\nu+\bar{\nu}$  & 2310 & 3111 & 1.35 & A2\\
     & $W >3.5$ GeV  & $\nu$           & 1258 & 1783 & 1.42 & A2$\nu$\\ 
     &                & $\bar{\nu}$     & 1052 & 1199 & 1.14 & A2$\bar{\nu}$\\ 
\hline
\hline
$\msbar$ & $Q >1.3$ GeV  & $\nu+\bar{\nu}$  & 2691 & 3732 & 1.39 & M\\
           & no $W_{cut}$       & $\nu$           & 1459 & 2205 & 1.51 & M$\nu$\\ 
     &                    & $\bar{\nu}$           & 1232 & 1419 & 1.15 & M$\bar{\nu}$\\ 
\hline
$\msbar$ & $Q >2$ GeV    & $\nu+\bar{\nu}$        & 2310 & 3080 & 1.33 & M2\\
     & $W >3.5$ GeV  & $\nu$                 &1258      & 1817  & 1.44   & M2$\nu$\\ 
     &                & $\bar{\nu}$     & 1052  & 1201 & 1.14 & M2$\bar{\nu}$\\ 
\hline
\end{tabular}
\end{center}
\caption{Fits to NuTeV cross section and dimuon data.}
\label{tab:Nutev}
\end{table}
%%%%%%%%%%%%%%%%%%%%%%%%%%%%%%%%%%%%%%%%%%%%%%%%%%%%%%%%%%%%%%%%%%%%%%%%%%%%%%%%%%%%%%%

The results of our fits to the NuTeV iron cross section and dimuon
data are summarized in Table~\ref{tab:Nutev}.
The cross section data have been corrected for QED radiation effects, 
and the non-isoscalarity of the iron target \cite{isoc}; correspondingly, 
we have used $A=56, Z=28$ in Eqs.~(\ref{eq:pdf}) and (\ref{eq:sfs}).\footnote{We have 
checked that omitting the isoscalar correction factors
and using $A=56, Z=26$ gives almost identical results.}
Note, for an iron target the isoscalar correction factors are small
and do not exceed the few \% level.
We have performed fits to the combined data as well as to the
neutrino- and anti-neutrino data sets separately.
Furthermore, two different cuts in the kinematic plane have been
examined: a) $Q > 1.3~\gev$, no cut on the hadronic invariant mass $W$
and b) $Q> 2~\gev$ and $W > 3.5~\gev$, {\it
cf.,}~Table~\ref{tab:Nutev}.
The NLO QCD calculation was performed in both the $\msbar$ and ACOT
schemes.  The ACOT scheme calculation takes into account the heavy
quark mass effects, whereas the $\msbar$ scheme assumes massless
partons.  The dominant target mass effects have been 
incorporated \cite{Kretzer:2003iu,Schienbein:2007gr}.\footnote{Target mass effects
(TMC) are expected to be relevant at large Bjorken-$x$ or small
momentum transfers $Q^2$ \protect\cite{Schienbein:2007gr}.
For issues of higher orders and higher twist {\it cf.} 
Refs.~\cite{Kataev:1997nc,Kataev:1998ce,Kataev:1999bp,Alekhin:1998df} 
}

As noted above, we have found bands for each class of fits from which
we have chosen central representatives.
The $\chi^2$ values have been determined taking into account the full
correlations of the data employing the effective $\chi^2$ function
given in Eq.~(B.5) of Ref.~\cite{Pumplin:2002vw}.
The numbers for the $\chi^2/pts$ are roughly on the order of $1.4$ for
both the ACOT and the $\msbar$ 
schemes.\footnote{Fits to this same
data neglecting the correlations between the errors and using the
conventional $\chi^2$ function ({\it cf.} Eq.~(B.1) in
\protect\cite{Pumplin:2002vw}), have  smaller $\chi^2/pts \simeq
1$.  While the uncorrelated errors are larger, the extracted
parameters are similar.}  
Furthermore, the fits to the anti-neutrino
data have considerably better $\chi^2$ values; however, we will see
below that this is at least partly due to the larger uncertainties of
these data.

%%%%%%%%%%%%%%%%%%%%%%%%%%%%%%%%%%%%%%%%%%%%%%%%%%%%%%%%%%%%%%%%%%%%%
% COMMENT ON BASE-1 AND BASE-2
%
\subsubsection{PDF Reference Sets}
For the purposes of this study, we use two different reference sets of
free-proton PDFs which we denote \hbox{`Base-1'} and \hbox{`Base-2'.}

Since we 
focus on iron PDFs and the associated nuclear corrections, we need a 
base set of PDFs which are essentially free of any nuclear effects;
this is the purpose of the  Base-1  reference set \cite{Owens:2007kp}. 
Therefore, to extract the Base-1 PDFs we omit the CCFR and NuTeV data 
from our fit 
so that our base PDFs do not contain any large residual nuclear
corrections.\footnote{%
We 
do retain the deuteron data as this has only
a small correction over the central $x$-range, ({\it cf.}
Sec.~\ref{sec:deuteron}) \cite{Gomez:1993ri,Owens:2007kp}.
The deuteron correction has been applied in the Base-1 fit. 
 Also, for the Drell-Yan Cu data (E605), the expected nuclear corrections in
this kinematic range are small (a few percent) compared to the overall
normalization uncertainty (15\%) and systematic error (10\%).  
}
The absence of such nuclear effects will be important 
in Sec.\ref{sec:nfac} when we extract the nuclear corrections factors.

The Base-2 PDFs are essentially the CTEQ6.1M PDFs with a 
modified  strange PDF  introduced to accommodate the NuTeV dimuon
data.\footnote{%
These PDFs have been determined from a fit to the same
data set as in the CTEQ6 analysis with the addition of the the NuTeV
dimuon data. The changes to the strange sea induce only minor changes to
the other fit parameters; this has a minimal effect 
on the particular observables ($d\sigma$, $F_2$) 
we examine in the present study. 
}
In the manner of the CTEQ6.1M PDF's,  the Base-2 fit does not 
apply any deuteron corrections to the data; this is in contrast to 
the Base-1 PDFs.
Also, the Base-2 fit does include the CCFR data that has been corrected to a
free nucleon using charged-lepton correction factors; the Fermilab
CCFR
experiment is the predecessor of NuTeV with comparable statistics as
those from NuTeV \cite{Yang:2001rm}.
The CCFR results in the large-$x$
region ($x>0.4$) are consistently lower than those from NuTeV, and
various sources contributing to the difference have been
identified \cite{Tzanov:2005kr,Tzanov:2005fu}. One third of the discrepancy has been
attributed to a mis-calibration of the magnetic field map of the muon
spectrometer, {\it i.e.}, to the muon energy scale in the CCFR
analysis.  About another third comes from model differences (cross
section model, muon and hadron energy smearing models).  A comparison
of NuTeV and CCFR data can be found in Ref.~\cite{Tzanov:2005kr}.

By comparing the free-proton PDF  \hbox{`Base-1'} and \hbox{`Base-2'}
sets with
the iron PDF sets of Table~\ref{tab:Nutev}, we can gauge the size of
the nuclear effects.  
Furthermore, differences between observables using the `Base-1'
respectively the `Base-2' reference sets will indicate the uncertainty
due to the choice of the free-proton PDF.\footnote{%
All results have been computed with both Base-1 and Base-2 PDFs.
Since the Base-2 PDFs use CCFR and NuTeV data, the resulting PDFs will
depend on the nuclear corrections which we are trying to determine.
Therefore,we will predominantly display the Base-1
PDFs for comparison in the following Sections.
}%%% END FOOTNOTE

%%%%%%%%%%%%%%%%%%%%%%%%%%%%%%%%%%%%%%%%%%%%%%%%%%%%%%%%%%%%%%%%%%%%%
% Figures 1-3:  A2 Fits 
\subsubsection{Comparison of the Fits with Data}

%%%%%%%%%%%%%%%%%%%%%%%%%%%%%%%%%%%%%%%%%%%%%%%%%%%%%%%%%%%%%%%%%%%%%
\figAAa
\figAAb
\figAAc
%\clearpage 

The quality of our fits of Table~\ref{tab:Nutev} can also be observed directly
in Figures~\ref{fig:fig2a} -- \ref{fig:fig2c} where we compare the
theoretical cross section $(1/E) d^2 \sigma/dx\, dy$ with a selection of
the data.
To be specific, we show all the data taken with beam energies $E =65,
150$, and $245~\gev$ which pass our kinematic cuts.  The measurements
are organized in bins of $x$ as a function of the inelasticity $y$ and
cover the $x$-range $0.015 \le x \le 0.750$.  The momentum transfers
can be computed using the relation $Q^2 = 2 M E x y$.
We normalize these plots using the `A2' fit which implements the
kinematic cuts $Q>2~\gev$ and $W> 3.5~\gev$ ({\it cf.}
Table~\ref{tab:Nutev}).  We note that these are the cuts employed in
the CTEQ6 analysis in order to reduce the sensitivity to target mass
and higher twist effects.\footnote{Conversely, global analyses of
nuclear PDFs tend to use looser kinematic cuts due to the lack of
small-$x$ data and the interest in the very large-$x$ region.}

The fit provides a good description of the data which are distributed
around unity for most of the bins.
For reference, the results of fit `A' (solid line) and Base-1 PDFs (dotted
line) are shown as well.
For fit `A2', the effect of the $Q>2~\gev$ cut is to remove
data at low $y$ in the small-$x$ region, and the $W>3.5~\gev$ cut
excludes low-$y$ data at large $x$.
The effects of these cuts on the fit are visible by comparing the
difference of the solid line (`A') from unity (`A2').
For $x\gsim 0.045$, we observe minimal differences
between the `A' and `A2' fits, and conclude the effect of the
kinematic cuts ($Q>2~\gev$ and $W> 3.5~\gev$) are nominal in this
region.
In the lowest $x$ bin ($x\sim 0.015$), 
much of the data is eliminated by the $Q>2~\gev$ cut such that
fit `A2' is only constrained
by a few data points at large $y$ for the higher neutrino energies, cf.~Fig.~\ref{fig:fig2c}.
Since both, fit `A' and fit 'A2', have large uncertainties 
in this $x$-region the comparison of individual representatives
is less significant---in particular at medium and low $y$ where no
data points lie.
In conclusion, we discern no relevant differences between the two 
classes of fits over the entire kinematic plane and will therefore
mainly focus on fit `A2' in the following sections.

%%%%%%%%%%%%%%%%%%%%%%%%%%%%%%%%%%%%%%%%%%%%%%%%%%%%%%%%%%%%%%%%%%%%
%  COMMENTS ON THE BASE PDFS IN FIGS 1-3 AND 4-6
%
\subsubsection{Comparison of the Fits with  Reference PDFs}

The dotted curve in Figures~\ref{fig:fig2a} -- \ref{fig:fig2c} shows
the cross sections obtained with Base-1 free-proton PDFs,
inserted into Eq.~\eqref{eq:pdf} to obtain ``free iron'' PDFs, divided
by the cross sections computed with fit `A2' PDFs.
The Base-2 PDFs (not shown) yield similar results as we demonstrate in 
Sec.~\ref{sec:xsec}. 
We expect the base PDFs will provide a poorer description of the data
since the nuclear modifications are not taken into account; the
deviations of these curves from unity indicate the size of the nuclear
effects.

We observe that the Base-1 results at small-$x$
($x\sim[0.045 -  0.08]$) are generally below unity (the `A2' fit)
 in the $y$ region of the data points
implying an enhancement due to nuclear effects.
As discussed above, the results in the lowest $x$ bin ($x=0.015$) are
less clear as the uncertainties are  larger since the kinematic cuts
remove much of the data. 
 Nevertheless,  do not see a clear signal of shadowing in this region
({\it cf.},  Fig.~\ref{fig:fig2c} at large $y$).

For intermediate $x\sim[0.125 - 0.175]$ the Base-1 (dotted line)
results are very similar to fit `A2'.  For larger $x\sim[0.225- 0.65]$
we observe a suppression of the nuclear cross sections qualitatively
similar to what is known from charged lepton DIS.
Finally, in the region $x\gsim 0.75$ the nuclear cross section is
again enhanced---an effect usually attributed to the Fermi motion of
the nucleons in the nucleus.

In conclusion, we observe the following pattern for the nuclear cross section
compared to the free nucleon cross section:
i) enhancement for $x \gsim 0.75$, ii) suppression for $x \sim [0.225 - 0.65]$,
iii) equality for $x \sim 0.125$, and iv) slight enhancement for $x \sim [0.045 - 0.08]$.
This is to be contrasted with the expectation from charged lepton DIS with the
well-known pattern:
i) enhancement for $x \gsim 0.75$ (Fermi motion), ii) suppression for
$x \sim [0.3 - 0.8]$ (EMC effect), iii) enhancement for $x \sim [0.06 - 0.3]$ 
(Anti-shadowing), and iv) suppression for $x \lsim 0.06$ (Shadowing).
Thus, for $x \gsim 0.3$ our results are generally as expected. However, we find 
 that the usual behavior at medium and small $x$ is modified.
We will examine this further in the following sections.

%%%%%%%%%%%%%%%%%%%%%%%%%%%%%%%%%%%%%%%%%%%%%%%%%%%%%%%%%%%%%%%%%%%%
\subsection{Iron PDFs}
%
% Figure 7: PDF PLOTS FOR ``A2'' FIT
%

%%%%%%%%%%%%%%%%%%%%%%%%%%%%%%%%%%%%%%%%%%%%%%%%%%%%%%%%%%%%%%%%%%%%
\figPDF

Having established the quality of our fits, 
we now examine the nuclear (iron) parton distributions $f_i^A(x,Q^2)$
according to Eq.~\eqref{eq:pdf}.  Figure~\ref{fig:fig3b} shows the
PDFs from fit `A2' at our input scale $Q_0=m_c=1.3~\gev$ versus $x$.
For an almost isoscalar nucleus like iron the $u$ and $d$
distributions are very similar, see Eqs.\ (\ref{eq:deviation1})--(\ref{eq:deviation4}).
Therefore, we only show the $u_v$ and
$\bar{u}$ partons, together with the strange sea.\footnote{While iron
is roughly isoscalar, other nuclear PDFs can exhibit larger
differences between the $u$ and $d$ distributions---the extreme case
being the free-proton PDF.  When comparing PDFs of
Eq.~\protect\eqref{eq:pdf}, we must keep in mind that it is ultimately
the structure functions defined by Eq.~\protect\eqref{eq:sfs2} which
are the physical observables.  }
As explained above, the gluon distribution is very similar to the
familiar CTEQ6M gluon at the input scale such that we don't show it
here.
In order to indicate the constraining power of the NuTeV data, the
band of reasonable fits is depicted.
The fits in this band were obtained (as outlined above) by varying the
initial conditions and the number of free parameters to fully explore
the solution space. All the fits shown in the band have $\chi^2/DOF$
within 0.02, which roughly corresponds to a range of $\Delta\chi^2\sim
50$ for the 2691 data points.

As can be seen in Figure~\ref{fig:fig3b}, the $u_v$ distribution
(Fig.~\ref{fig:fig3b}a) has a very narrow band across the entire
$x$-range.
The up- and strange-sea distributions (Fig.~\ref{fig:fig3b}b and
Fig.~\ref{fig:fig3b}c) are less precisely determined.  At values of $x$
down to, say, $x\simeq 0.07$ the bands are still reasonably well
confined; however, they open up widely in the small-$x$ region.
Cases where the strange quark sea lies above the up-quark sea are
unrealistic, but are present in some of the fits since this region ($x
\lesssim 0.02$) is not constrained by data.
We have included the curves for our free-proton Base-1 PDFs (dashed),
as well as 
the HKN04 \cite{Hirai:2004wq} (dotted), 
the NLO HKN07 \cite{Hirai:2007sx} (dotted-dashed), 
and DS \cite{deFlorian:2003qf} (dot-dashed) 
nuclear PDFs.\footnote{In a recent publication, Eskola
{\it et al.} \cite{Eskola:2007my} perform a global reanalysis of their
ESK98 \cite{Eskola:1998df} nuclear PDFs. While we do not present a
comparison here, the results are compatible with those distributions
displayed in Fig.~\ref{fig:fig3b}; a comparison
can be found in Figs.~10 and 11 of Ref.~\cite{Eskola:2007my}.  }

% The comparison with the Base-1 PDFs is straightforward since, with
% exception of the heavy flavor scheme (ACOT {\it vs.} $\msbar$), the same
% theoretical framework (input scale, functional form, NLO evolution)
% has been utilized for their determination.
%
The comparison with the Base-1 PDFs is straightforward since 
 the same theoretical
framework (input scale, functional form, NLO evolution) has been
utilized for their determination.
Therefore, the differences between the solid band and the dashed line
exhibit the nuclear effects, keeping in mind that the free-proton PDFs
themselves have uncertainties.

For the comparison with the HKN04 distributions, it should be noted that
a SU(3)-flavor symmetric sea has been used; therefore, the HKN04 strange
quark distribution is larger, and the light quark sea smaller, than
their Base-1 PDF counterparts over a wide range in $x$.  Furthermore,
the HKN04 PDFs are evolved at leading order.

In a recent analysis,  the HKN group has published a new set of 
NPDFs (HKN07) including uncertainties \cite{Hirai:2007sx}. 
They provide  both LO and NLO sets of PDFs, and we display 
the NLO set. These PDFs  also use a more general 
set of sea distributions such that $\bar{u}(x)\not=\bar{d}(x)\not=\bar{s}(x)$
in general.

The DS PDFs are linked to the GRV98 PDFs \cite{Gluck:1998xa} with a
rather small radiatively generated strange sea distribution.
Consequently, the light quark sea is enhanced compared to the other
sets.
Additionally, the DS sets are evolved in a 3-fixed-flavor scheme in
which no charm parton is included in the evolution. However, at the
scale $Q=m_c$ of Fig.~\ref{fig:fig3b} this is of no importance.

%%%%%%%%%%%%%%%%%%%%%%%%%%%%%%%%%%%%%%%%%%%%%%%%%%%%%%%%%%%%%%%%%%%
\addtocontents{toc}{\protect  \newpage}
\addtocontents{toc}{\protect \null}
\addtocontents{toc}{\protect  \vspace{2mm}}
\section{Nuclear Correction Factors \label{sec:nfac}}
In the previous section we analyzed
charged current $\nu$--Fe data with the goal of extracting the  iron
nuclear parton distribution
functions.
In this section, we now compare our iron PDFs with the free-proton PDFs 
(appropriately scaled) to infer the proper heavy target correction which 
should be applied to relate these  quantities.

Within the parton model, a nuclear correction factor $R[{\cal O}]$
for an observable ${\cal O}$ can be defined as follows: 
\begin{equation}
R[{\cal O}] = \frac{{\cal O}[{\rm NPDF}]}{{\cal O}[{\rm free}]}
\label{eq:R}
\end{equation}
where ${\cal O}[{\rm NPDF}]$ represents the observable computed with nuclear PDFs, 
and ${\cal O}[{\rm free}]$ is the same observable 
constructed out of the free nucleon PDFs according to
Eq.~\eqref{eq:pdf2}.
Clearly, $R$ can depend on the observable under consideration
simply because different observables may be sensitive to different 
combinations of PDFs.

This means that the nuclear correction factor $R$ 
for 
$F_2^A$ and 
$F_3^A$ 
will, in general,  be different.  Additionally, the nuclear correction factor for
$F_2^A$ will yield different results for the  charged current
$\nu$--$Fe$ process ($W^\pm$ exchange) as compared with the neutral
current $\ell^\pm$--$Fe$ process ($\gamma$ exchange).
Schematically, we can write the nuclear correction for the DIS structure function
$F_2$ in a charged current (CC) $\nu$--$A$ process as (cf.\ Eq.\ (\ref{eq:f2nu})):\footnote{The
corresponding anti-neutrino process is obtained with a
$u\leftrightarrow d$ interchange.}
%%%%%%%%%%%%%%%%%%%%%%%%%%
\def\p{{\,\emptyset }}
%%%%%%%%%%%%%%%%%%%%%%%%%%
\begin{eqnarray}
R_{CC}^{\nu} (F_2;x,Q^2)
&\simeq&
\frac{d^A + \bar{u}^A + ...}{d^\p + \bar{u}^\p + ... }
\label{eq:rcc}
\end{eqnarray}
and contrast this with the neutral current (NC) 
$\ell^\pm$--$A$ process (cf.\ Eq.\ (\ref{eq:f2em})):
\begin{eqnarray}
&&R_{NC}^{e,\mu} (F_2;x,Q^2)
\simeq
\nonumber \\[5pt]
&&\qquad
\frac{ 
 \left(-\frac{1}{3}\right)^2  \left[ d^A + \bar{d}^A + ...\right]
+\left(+\frac{2}{3}\right)^2  \left[ u^A + \bar{u}^A + ...\right]
}{
 \left(-\frac{1}{3}\right)^2  \left[ d^\p + \bar{d}^\p + ...\right]
+\left(+\frac{2}{3}\right)^2  \left[ u^\p + \bar{u}^\p + ...\right]
 }
\label{eq:rnc}
\ ,
\nonumber \\
\end{eqnarray}
 where the superscript ``$\p$'' denotes the ``free nucleon'' PDF
which is constructed via the relation: 
\begin{equation}
f_i^\p(x,Q) = \frac{Z}{A}\  f_i^{p}(x,Q) +  \frac{(A-Z)}{A}\  f_i^{n}(x,Q)
\quad .
\label{eq:pdf2}
\end{equation}
Clearly, the $R$-factors depend on both 
the kinematic variables and the factorization scale.
Finally, we note that
Eq.~\eqref{eq:R} is subject to uncertainties of both the numerator and 
the denominator.

We will now evaluate the nuclear correction factors for our extracted
PDFs, and compare these with selected results from the 
literature \cite{Kulagin:2004ie,Kulagin:2007ju,NMC}.\footnote{Note that our
comparison with the Kulagin--Petti model is based on the work in
Ref.~\cite{Kulagin:2004ie}.}
Because we have extracted the iron PDFs from only iron data, we do
not assume any particular form for the nuclear $A$-dependence; hence
the extracted $R[{\cal O}]$ ratio is essentially model independent.

%%%%%%%%%%%%%%%%%%%%%%%%%%%%%%%%%%%%%%%%%%%%%%%%%%%%%%
\subsection{Deuteron corrections for the $F_2^{Fe}/F_2^D$ ratio
\label{sec:deuteron}}
%

%%%%%%%%%%%%%%%%%%%%%%%%%%%%%%%%%%%%%%%%%%%%%%%%%%%%%%%%%%%%%%%%%%%%
\figDeuteron

The structure function ratio $F_2^{\rm Fe}/F_2^{\rm D}$ provides a
common (and
useful) observable to use to gauge the nuclear effects of iron.
To construct  the
numerator, we will use our iron PDFs as extracted in fits `A' and `A2.'
For the denominator, we will use the Base-1 and Base-2 free proton
PDF; however, converting from free proton structure functions to
%However, this figure portrays an unrealistic situation at large $x$; ...
deuteron structure functions is nontrivial and model-dependent.

In Fig.~\ref{fig:fig4} we display the NMC data for $F_2^D / F_2^p$
\cite{Arneodo:1996kd} and compare this to a variety of data
parameterizations \cite{Arneodo:1996kd,Gomez:1993ri,Owens:2007kp,Tvaskis:2004qm,Tvaskis:2006tv}.
The dashed line shows the structure function ratio computed with the
Base-1 PDFs; in this case a nuclear correction factor for deuterium
has been applied using the parameterization of
Ref.~\cite{Owens:2007kp}.
The solid line shows the structure function ratio computed with the
Base-2 PDFs; in this case no nuclear correction factor for deuterium
was applied.
The dotted line (Arneodo) is the parameterization of Ref.~\cite{Arneodo:1996kd},
and the dot-dashed line (Tvaskis) is the parameterization of
Ref.~\cite{Tvaskis:2004qm,Tvaskis:2006tv}. 
We see that the range of discrepancies in the deuterium corrections
 are typically on
the order of a percent or two except at large $x$; while this
correction cannot be neglected, it is small compared to the much
larger iron nuclear corrections. To explore a range of possibilities
(reflecting the underlying uncertainty) we have incorporated deuteron
corrections into the Base-1 PDF, but not the Base-2 PDF; hence the
spread between these two reference PDFs will, in part, reflect our
ignorance of $F_2^D$ and other uncertainties of proton PDFs at
large-$x$.

%%%%%%%%%%%%%%%%%%%%%%%%%%%%%%%%%%%%%%%%%%%%%%%%%%%%%%
\subsection{$F_2^{Fe}/F_2^D$ for neutral current (NC) charged lepton scattering}

%%%%%%%%%%%%%%%%%%%%%%%%%%%%%%%%%%%%%%%%%%%%%%%%%%%%%%%%%%%%%%%%%%%%
\figSLAC

We will also find it instructive to compare our results with the
$F_2^{\rm Fe}/F_2^{\rm D}$ as extracted in neutral current
charged-lepton scattering, $\ell^\pm$--$Fe$.
In Fig.~\ref{fig:fig5} we compare  the experimental results for the structure
function ratio $F_2^{\rm Fe}/F_2^{\rm D}$  for the following experiments: 
\hbox{BCDMS-85} \cite{Bari:1985ga}, 
\hbox{BCDMS-87} \cite{Benvenuti:1987az}, 
\hbox{SLAC-E049} \cite{Bodek:1983qn}, 
\hbox{SLAC-E139} \cite{Gomez:1993ri}, 
\hbox{SLAC-140} \cite{Dasu:1993vk}.
The curve (labeled SLAC/NMC parameterization) is a fit to this data \cite{NMC}.
While there is a spread in the individual data points, the
parameterization describes the bulk of the data at the level of a few
percent or better.  It is important to note that this parameterization
is independent of atomic number $A$ and the energy scale $Q^2$ \cite{Arrington:2003nt};
this
is in contrast to the results we will derive using the PDFs extracted
from the nuclear data.\footnote{In particular, we will find for large
$x$ ($\gsim 0.5$) and $Q$ comparable to the proton mass the target
mass corrections for $F_2^{\rm Fe}/F_2^{\rm D}$ are essential for
reproducing the features of the data; hence the $Q$ dependence plays
a fundamental role.} 
Additionally, we note that while this parameterization has been extracted
using ratios of $F_2$ structure functions, it is often applied to other
observables such as $F_{1,3,L}$ or $d\sigma$. 
 We can use this parameterization as a guide to
judge the approximate correspondence between this neutral current (NC)
charged lepton DIS data and our charged current (CC) neutrino DIS
data.

%%%%%%%%%%%%%%%%%%%%%%%%%%%%%%%%%%%%%%%%%%%%
%FIGURE RATIO XSEC
\subsection{Correction Factors for $d^2\sigma/dx \, dQ^2$
\label{sec:xsec}}

%%%%%%%%%%%%%%%%%%%%%%%%%%%%%%%%%%%%%%%%%%%%%%%%%%%%%%%%%%%%%%%%%%%%
\figDSIG
We begin by computing the nuclear correction factor $R$ 
according to Eq.\ (\ref{eq:R}) for the neutrino differential cross section 
in Eq.\ (\ref{eq:dsig}) as this represents
the bulk of the NuTeV data included in our fit.
More precisely, we show $R$-factors for the charged current cross sections
$d^2\sigma/dx \, dQ^2$ at fixed $Q^2$ which can be obtained from 
Eq.\ (\ref{eq:dsig}) by a simple Jacobian transformation and we consider
an iron target which has been corrected for the neutron excess, i.e.,
we use the PDFs in Eq.\ (\ref{eq:pdf}) (for the numerator) 
and Eq.\ (\ref{eq:pdf2}) (for the denominator)
with $A=56$ and $Z=28$.
Our results are displayed in Fig.~\ref{fig:fig8} for $Q^2=5~\gevsq$
and a neutrino energy $E_\nu = 150\ \gev$ which implies,
due to the relation $Q^2 = 2 M E_\nu x y$,
a minimal $x$-value of $x_{{\rm min}} = 0.018$.
The solid (dashed) lines correspond to neutrino (anti-neutrino)
scattering using the iron PDFs from the `A2' fit.

We have computed $R$ using both the Base-1 and Base-2 PDFs for the
denominator of Eq.~(\ref{eq:R}); recall that Base-1 
includes a deuteron correction while Base-2 uses the CCFR data and
does not include a deuteron correction.
The difference between the Base-1 and Base-2 curves is approximately
2\% at small $x$ and grows to 5\% at larger $x$, with Base-2 above the
Base-1 results. 
As this behavior is typical, in the following plots  
(Figs.~\ref{fig:fig6a} and Figs.~\ref{fig:fig6b}) we
will only show the Base-1 results.
We also observe that the neutrino (anti-neutrino) results coincide in
the region of large $x$ where the valence PDFs are dominant, but
differ by a few percent at small $x$ due to the differing strange and
charm distributions.

%%%%%%%%%%%%%%%%%%%%%%%%%%%%%%%%%%%%%%%%%%%%
%FIGURE RATIO NEUTRINO
\subsection{Correction Factors for  $F_2^\nu(x,Q^2)$ and  $F_2^{\bar\nu}(x,Q^2)$}

%%%%%%%%%%%%%%%%%%%%%%%%%%%%%%%%%%%%%%%%%%%%%%%%%%%%%%%%%%%%%%%%%%%%
\figNU
\figNUBAR

We now compute the nuclear correction factors for charged current
 neutrino--iron scattering.  The results for $\nu$--$Fe$ are shown in
 Fig.~\ref{fig:fig6a}, and those of $\bar\nu$--$Fe$ are shown in
 Fig.~\ref{fig:fig6b}.
The numerator in Eq.~\eqref{eq:R} has been computed using the nuclear
PDF from fit `A2', 
and for the denominator 
we have used  the Base-1 PDFs.
For comparison we also show the correction factor from the Kulagin--Petti model
\cite{Kulagin:2004ie} (dashed-dotted), and 
the SLAC/NMC curve (dashed) \cite{NMC}
which has been obtained from an $A$ and $Q^2$-independent parameterization 
of calcium and iron charged--lepton DIS data.

Due to the neutron excess in iron,\footnote{Note that the correction
factors shown in Figs.\ \protect\ref{fig:fig6a} and \protect\ref{fig:fig6b}
are valid for the case in which the data have {\em not} been corrected for
the neutron excess in iron.
For data that already have been corrected for the neutron excess one should, 
for consistency, compute the $R$-factors using $A=56$, $Z=28$ in
equation Eq.~\eqref{eq:pdf}. 
The magnitude of the difference between the $R$-factors in these two cases ($Z=26$ vs.\ $Z=28$)
is typically a few percent.} 
both our curves and the KP curves differ when comparing scattering for
neutrinos (Fig.~\ref{fig:fig6a}) and anti-neutrinos
(Fig.~\ref{fig:fig6b}); the SLAC/NMC parameterization is the same in
both figures.
For our results (solid lines), the difference between the neutrino and
anti-neutrino results is relatively small, of order $3 \%$ at $x=0.6$.
Conversely, for the KP model (dashed-dotted lines) the
$\nu$--$\bar\nu$ difference reaches $10 \%$ at $x\sim 0.7$, and
remains sizable at lower values of $x$.

To demonstrate the dependence of the $R$ factor on the kinematic
variables, in Figs.~\ref{fig:fig6a} and Fig.~\ref{fig:fig6b} we have
plotted the nuclear correction factor for two separate values of
$Q^2$.  Again, our curves and the KP model yield different results for
different $Q^2$ values, in contrast to the SLAC/NMC parameterization.

Comparing the nuclear correction factors for the $F_2$ structure
function (Figs.~\ref{fig:fig6a} and Fig.~\ref{fig:fig6b}) with those
obtained for the differential cross section (Fig.~\ref{fig:fig8}), we
see these are quite different, particularly at small $x$. Again, this
is because the cross section $d^2\sigma$ is comprised of a
different combination of PDFs than the $F_2$ structure function.  In
general, our $R$-values for $F_2$ lie below those of the corresponding
$R$-values for the cross section $d\sigma$ at small $x$. 
Since $d\sigma$ is a
linear combination of $F_2$ and $F_3$, the $R$-values for $F_3$ (not
shown) therefore lie above those of $F_2$ and $d\sigma$. 
Again, we
emphasize that it is important to use an appropriate nuclear
correction factor which is matched to the particular observable.

As we observed in the previous section, our results have general
features in common with the KP model and the SLAC/NMC
parameterization, but the magnitude of the effects 
and the $x$-region where they apply
are quite
different.  
Our results are noticeably flatter than the KP and
SLAC/NMC curves, especially at moderate-$x$ where the differences are significant.  
The general trend we see when examining these nuclear correction
factors is that the anti-shadowing region is shifted to smaller $x$
values and any turn-over at low $x$ is minimal given the PDF
uncertainties.
In general, these plots
suggest that the size of the nuclear corrections extracted from the
NuTeV data are smaller than those obtained from charged lepton
scattering (SLAC/NMC) or from the set of data used in the KP model.
We will investigate this difference further in the following section.

%%%%%%%%%%%%%%%%%%%%%%%%%%%%%%%%%%%%%%%%%%%%
%\subsection{Correction Factors for Charged-Lepton $F_2^{Fe}/F_2^D$}
\subsection{Predictions for Charged-Lepton $F_2^{Fe}/F_2^D$ from iron PDFs}

%%%%%%%%%%%%%%%%%%%%%%%%%%%%%%%%%%%%%%%%%%%%%%%%%%%%%%%%%%%%%%%%%%%%
\figFINAL

Since the SLAC/NMC parameterization was fit to $F_2^{Fe}/F_2^D$ for
charged-lepton DIS data, we can perform a more balanced comparison by
using our iron PDFs to compute this same quantity. The results are
shown in Fig.~\ref{fig:ccncCompare} where
we have used our iron PDFs to
compute $F_2^{Fe}$, and the Base-1 and Base-2 PDFs to compute
$F_2^{D}$.

As with the nuclear correction factor results of the previous section, we 
find our results  have some gross features in common 
while on a more refined level the magnitude of the  nuclear
corrections extracted from the CC iron data differs from the 
charged lepton  data.  
In particular, we note that the so-called ``anti-shadowing''
enhancement at $x \sim [0.06-0.3]$ is {\it not} reproduced by the charged
current (anti-)neutrino data.  
Examining our results among all the various $R[{\cal O}]$
calculations, we generally find that any nuclear enhancement in the
small $x$ region is reduced and shifted to a lower $x$ range as
compared with the SLAC/NMC parameterization.
In fact, this behavior is expected given the comparisons of
Figs.~\ref{fig:fig2a}--\ref{fig:fig2c} which show that at $x\sim 0.1$
the cross sections obtained with the base PDFs are not smaller than
the `A' and `A2' fitted cross sections.
Furthermore, in the limit of large $x$ ($x \gtrsim 0.6$) our results
are slightly higher than the data, including the very precise
SLAC-E139 points; however,the large theoretical uncertainties on $F_2^D$ in this
$x$-region (see Fig.~\ref{fig:fig4}) make it difficult to extract firm conclusions.

This discussion raises the more general question as to whether the
charged current ($\nu$--$Fe$) and neutral current ($\ell^\pm$--$Fe$)
correction factors are entirely
compatible 
\cite{Tzanov:2005kr,Tzanov:2005fu,Boros:1998qh,Boros:1999fy,Bodek:1999bb,Kretzer:2001mb}. There is {\it a priori}
no requirement that these be equal; in fact, given that the $\nu$--$Fe$
process involves the exchange of a $W$ and the $\ell^\pm$--$Fe$ process
involves the exchange of a $\gamma$ we necessarily expect this will
lead to differences at some level.
To say definitively how much of this difference is due to this effect
and how much is due to the uncertainty of our nuclear PDFs requires
further study; in particular, it would be interesting to extend the
global analysis of nuclear PDFs to include neutral current
charged-lepton as well as additional charged current neutrino data.
Here, the analysis of additional data sets such as the ones from the CHORUS
experiment \cite{Onengut:2005kv,KayisTopaksu:2003mt} (neutrino-lead
interactions) should help clarify these questions.
We are in the processes of adding additional nuclear data sets to our
analysis; however, this increased precision comes at the expense of
introducing the ``A'' degree of freedom into the fit.

%%%%%%%%%%%%%%%%%%%%%%%%%%%%%%%%%%%%%%%%%%%%%%%%%%%%%%%%%5
\section{Conclusions}
\label{sec:summary}

We have presented a detailed analysis of the high statistics NuTeV
neutrino--iron data in the framework of the parton model at
next-to-leading order QCD.
This investigation takes a new approach to this problem by studying a
single nuclear target (iron) so that we avoid the difficulty of having
to assume a nuclear ``A''-dependence.
In this context, we have extracted a set of iron PDFs which are free
of any nuclear model dependence.
By comparing these iron PDFs with ``free proton'' PDFs, we can
construct the associated nuclear correction factor $R$ for any chosen
observable in any given $\{x,Q^2\}$ kinematic range.

While the nuclear corrections extracted from charged current
$\nu$--$Fe$ scattering have similar characteristics as the neutral
current $l^\pm$--$Fe$ charged-lepton results, the detailed $x$ and $Q^2$
behavior is quite different.
These results raise the deeper question as to whether the charged
current and neutral current correction factors may be substantially
different.
A combined analysis of neutrino and charged-lepton data sets, for
which the present study provides a foundation, will shed more light on
these issues.
Resolving these questions is essential if we are to reliably use the
plethora of nuclear data to obtaining free-proton PDFs.

%%%%%%%%%%%%%%%%%%%%%%%%%%%%%%%%%%%%%%%%%%%%%%%%%%%%%%%%%%%%%%%%%%%%%%%%%%%%%%%

%%%%%%%%%%%%%%%%%%%%%%%%%%%%%%%%%%%%%%%%%%%%%%%%%%%%%%%%%%%%%%%%%%%%%%%
 
\section*{Acknowledgment}

We thank 
Tim Bolton,
Javier Gomez, 
Shunzo Kumano,
Eric Laenen,
Dave Mason, 
W.~Melnitchouk, 
Donna Naples,
Mary Hall Reno,
Voica~A.~Radescu,
and
Martin Tzanov
for valuable discussions.
F.I.O., I.S.,  and J.Y.Y.\  acknowledge the hospitality of Argonne,
BNL, CERN, and Fermilab  where a portion of this work was performed.  
%%%
This work was partially supported by the U.S.\ Department of Energy
under grant DE-FG02-04ER41299,  
contract DE-AC05-06OR23177 (under which Jefferson
Science Associates LLC operates the Thomas Jefferson National
Accelerator Facility), 
the National Science Foundation grant 0400332,
the
Lightner-Sams Foundation, 
and
the Sam Taylor Foundation. 
%%%
The work of J.F.~Owens was supported in part by the U.S.\ Department of
Energy under contract number DE-FG02-97IR41022.
The work of J.~Y.~Yu was supported by the
Deutsche Forschungsgemeinschaft (DFG) through grant No.~YU~118/1-1.

%%%%%%%%%%%%%%%%%%%%%%%%%%%%%%%%%%%%%%%%%%%%%%%%%%%%%%%%%%%%%%%%%%%%%%%
%%%%%%%%%%%%%%%%%%%%%%%%%%%%%%%%%%%%%%%%%%%%%%%%%%%%%%%%%%%%%%%%%%%%%%%

%%%%%%%%%%%%%%%%%%%%%%%%%%%%%%%%%%%%%%%%%%%%%%%%%%%%%%%%%%%%%%%%%%%%%%%%%%%%%%%%%%%%%%
%\newpage{}
%\newpage{}
%\section*{References}
%\bibliographystyle{unsrt}
%\bibliographystyle{JHEP}
\bibliographystyle{apsrev}
\bibliography{iron}
%%%%%%%%%%%%%%%%%%%%%%%%%%%%%%%%%%%%%%%%%%%%%%%%%%%%%%%%%%%%%%%%%%%%%%%%%%%%%%%%%%%%%%
\end{document}